%% file: main.tex
\documentclass[preprint]{aastex631}
\input{preamble}

\begin{document}

\title{Internal 1000 AU-scale Structures of the R CrA Cluster-forming Cloud\\
--- I: Filamentary Structures ---}

\author[0000-0002-1411-5410]{Kengo Tachihara}
\affiliation{Department of Physics, Graduate School of Science, Nagoya University, Furo-cho, Chikusa-ku, Nagoya 464-8601, Japan}

\author[0009-0005-4458-2908]{Naofumi Fukaya}
\affiliation{Department of Physics, Graduate School of Science, Nagoya University, Furo-cho, Chikusa-ku, Nagoya 464-8601, Japan}

\author[0000-0002-2062-1600]{Kazuki Tokuda}
\affiliation{Department of Earth and Planetary Sciences, Faculty of Science, Kyushu University, 744 Motooka, Nishi-ku, Fukuoka 819-0395, Japan}
\affiliation{National Astronomical Observatory of Japan, National Institutes of Natural Science, 2-21-1 Osawa, Mitaka, Tokyo 181-8588, Japan}

\author[0000-0001-5859-8840]{Yasumasa Yamasaki}
\affiliation{Department of Physics, Graduate School of Science, Osaka Metropolitan University, 1-1 Gakuen-cho, Naka-ku, Sakai, Osaka 599-8531, Japan}
\affiliation{National Astronomical Observatory of Japan, National Institutes of Natural Science, 2-21-1 Osawa, Mitaka, Tokyo 181-8588, Japan}

\author{Takeru Nishioka}
\affiliation{Department of Physics, Graduate School of Science, Nagoya University, Furo-cho, Chikusa-ku, Nagoya 464-8601, Japan}

\author[0000-0001-6891-2995]{Daisei Abe}
\affiliation{Department of Physics, Graduate School of Science, Nagoya University, Furo-cho, Chikusa-ku, Nagoya 464-8601, Japan}
\affiliation{Astronomical Institute, Tohoku University, Sendai, Miyagi, 980-8578, Japan}

\author[0000-0002-7935-8771]{Tsuyoshi Inoue}
\affiliation{Department of Physics, Konan University, Okamoto 8-9-1, Higashinada-ku, Kobe, Hyogo 658-8501, Japan}

\author[0000-0002-8217-7509]{Naoto Harada}
\affiliation{Department of Earth and Planetary Sciences, Graduate School of Science, Kyushu University, 744 Motooka, Nishi-ku, Fukuoka 819-0395, Japan}
\affiliation{Department of Astronomy, Graduate School of Science, The University of Tokyo,
7-3-1 Hongo, Bunkyo-ku, Tokyo 113-0033, Japan}

\author[0000-0001-6580-6038]{Ayumu Shoshi}
\affiliation{Department of Earth and Planetary Sciences, Graduate School of Science, Kyushu University, 744 Motooka, Nishi-ku, Fukuoka 819-0395, Japan}

\author[0000-0003-4271-4901]{Shingo Nozaki}
\affiliation{Department of Earth and Planetary Sciences, Graduate School of Science, Kyushu University, 744 Motooka, Nishi-ku, Fukuoka 819-0395, Japan}

\author[0000-0001-5817-6250]{Asako Sato}
\affiliation{Department of Earth and Planetary Sciences, Graduate School of Science, Kyushu University, 744 Motooka, Nishi-ku, Fukuoka 819-0395, Japan}

\author[0000-0002-7951-1641]{Mitsuki Omura}
\affiliation{Department of Earth and Planetary Sciences, Graduate School of Science, Kyushu University, 744 Motooka, Nishi-ku, Fukuoka 819-0395, Japan}

\author{Kakeru Fujishiro}
\affiliation{Department of Physics, Graduate School of Science, Nagoya University, Furo-cho, Chikusa-ku, Nagoya 464-8601, Japan}

\author[0000-0003-1117-9213]{Misato Fukagawa}
\affiliation{National Astronomical Observatory of Japan, National Institutes of Natural Science, 2-21-1 Osawa, Mitaka, Tokyo 181-8588, Japan}

\author[0000-0002-0963-0872]{Masahiro N. Machida}
\affiliation{Department of Earth and Planetary Sciences, Faculty of Science, Kyushu University, 744 Motooka, Nishi-ku, Fukuoka 819-0395, Japan}

\author{Takahiro Kanai}
\affiliation{Graduate School of Science and Engineering, Saitama University, 255 Shimo-Okubo, Sakura-ku, Saitama 338-8570, Japan}

\author[0000-0001-7249-6787]{Yumiko Oasa}
\affiliation{Faculty of Education/Graduate School of Science and Engineering, Saitama University, 255 Shimo-Okubo, Sakura-ku, Saitama 338-8570, Japan}

\author[0000-0001-7826-3837]{Toshikazu Onishi}
\affiliation{Department of Physics, Graduate School of Science, Osaka Metropolitan University, 1-1 Gakuen-cho, Naka-ku, Sakai, Osaka 599-8531, Japan}

\author[0000-0003-1549-6435]{Kazuya Saigo}
\affiliation{Graduate School of Science and Engineering, Kagoshima University, 1-21-40 Korimoto Kagoshima-city Kagoshima, 890-0065, Japan}

\author[0000-0002-8966-9856]{Yasuo Fukui}
\affiliation{Department of Physics, Graduate School of Science, Nagoya University, Furo-cho, Chikusa-ku, Nagoya 464-8601, Japan}

\keywords{Interstellar medium (847) --- Interstellar clouds (834) --- Interstellar filaments (842) --- Star formation (1569) --- Star clusters (1567)}

\input{0_Abstract}
\input{1_Introduction}
\input{2_Observations}
\input{3_Results}
\input{4_Discussions}

\input{5_Conclusions}
\input{6_Appendix}

\bibliography{RCrA_reference}
\bibliographystyle{aasjournal}

\end{document}

%% file: preamble.tex
\usepackage{amsmath}
\usepackage{savesym}
\savesymbol{tablenum}
\usepackage{siunitx}
\restoresymbol{SIX}{tablenum}

%% file: 0_Abstract.tex
\begin{abstract}

We report on ALMA ACA observations of a high-density region of the Corona Australis cloud forming a young star cluster, and the results of resolving internal structures. In addition to embedded Class 0/I protostars in continuum, a number of complex dense filamentary structures are detected in the C$^{18}$O and SO lines by the 7m array. These are sub-structures of the molecular clump that are detected by the TP array as the extended emission. We identify 101 and 37 filamentary structures with a few thousand AU widths in C$^{18}$O and SO, respectively, called as {\it feathers}. The typical column density of the feathers in C$^{18}$O is about $10^{22}$ cm$^{-2}$, and the volume density and line mass are $\sim 10^5$ cm$^{-3}$, and a few $\times M_{\sun}$ pc$^{-1}$, respectively. This line mass is significantly smaller than the critical line mass expected for cold and dense gas. These structures have complex velocity fields, indicating a turbulent internal property. The number of feathers associated with Class 0/I protostars is only $\sim 10$, indicating that most of them do not form stars but rather being transient structures. The formation of feathers can be interpreted as a result of colliding gas flow as the morphology well reproduced by MHD simulations, supported by the the presence of \ion{H}{1} shells in the vicinity. The colliding gas flows may accumulate gas and form filaments and feathers, and trigger the active star formation of the R CrA cluster.
\end{abstract}

%% file: 1_Introduction.tex
\section{Introduction}
    \label{fig:intro}

Star formation is the most fundamental and important physical process in galaxy evolution. To investigate the initial conditions for star formation, both observational and theoretical studies have focused on how dense molecular gas forms, fragments, and undergoes gravitational collapse. On the other hand, the distribution of young stars indicates that there are two modes of star formation: isolated star formation and clustered one, typically characterized by stellar densities of a few pc$^{-2}$ and a few hundred pc$^{-2}$, respectively \citep[e.g.,][]{carpenter2000}. For example in Taurus molecular cloud, only low-mass stars are born in isolation, while massive stars are formed as members of clusters in giant molecular clouds such as the Orion cloud complex \citep[e.g.,][]{larson1982}. There are various scales of star clusters, and the maximum mass of the member stars is supposed to be determined by the total mass of the system and the initial mass function (IMF), while \citet{bressert2010} claimed that they are no bimodal but continuous. It is still the subject of much debate whether these are formed by the same physical processes or not \citep[e.g.,][]{oasa2008}.

Focusing on the structure of the molecular clouds that are the progenitors of young stars, filamentary molecular clouds have been observed ubiquitously regardless of the star formation activities \citep[e.g.,][]{arzoumanian11, andre14} across the star-forming galaxies in the Local Group \citep[e.g.,][]{Tokuda19N159,Muraoka20,Neelamkodan21}. In relatively simple systems of low-mass star formation regions, filaments with 0.1 pc widths fragment along their long axis to form molecular cloud cores, where protostars are formed at the central dense parts. Theoretical studies have also shown that self-gravitating unstable filaments whose line masses is equal to the critical line mass fragment at a wavelength of about four times their radius, to form a few $M_{\sun}$ dense cores, and when their line masses exceed it, they collapse toward the axis and form stars \citep{ostriker64, inutsuka97}. For the filament formation, several models have been proposed, but still many unanswered issues remain such as why filaments with different properties have almost uniform widths of $\sim 0.1$ pc.

On the other hand, in young massive star-forming regions such as the InfraRed Dark Clouds (IRDCs), so-called hub-filament structures with multiple filaments converge at a single point are often observed, with young massive stars associated with the dense hub regions \citep[e.g.][]{myers09,Fukui19,Tokuda19N159}. Based on the velocity and magnetic field structures, collisions of filaments and channel gas flow along the filaments have been proposed as models for their formation \citep{liu2012, kirk2013, peretto14, nakamura2014, ren2021}. However, how filaments fragment and make individual star formation happen as clustered star formation remains to be elucidated. In particular, resolving filamentary structures and identifying embedded dense molecular cloud cores requires observations with high resolutions and large dynamic ranges in column density. Because of the large distances to typical cluster-forming regions, observational difficulties have been an obstacle to these studies.

The Corona Australis molecular cloud is one of the nearest star-forming regions at the distance of 149 pc \citep{galli2020} showing cluster formation containing three Herbig Ae/Be stars, R CrA, T CrA, and TY CrA \citep{herbst99}, with approximately 50 young stars concentrated in an area of 0.5 pc$^2$ as a Coronet cluster (aka R CrA cluster) reported by the ground-based infrared observations \citep{sicilia-aguilar11}, Spitzer Gould Belt Survey \citep{peterson11}, and Herschel Space Telescope \citep{sicilia-aguilar13}. The cluster is associated with the reflection nebula NGC 6729 \citep[see][for review]{neuhauser08}, and several embedded Class 0/I objects have been discovered, indicating ongoing active star formation \citep{groppi2004, hamaguchi2005, nutter2005, groppi07, peterson11, sicilia-aguilar13}. TY CrA and HD 176386 with reflection nebulae NGC 6726/6727, are located approximately 5 arcmin northwest of the R CrA cluster, while there is no noticeable concentration of embedded protostars around them \citep{peterson11}. In addition, a deep near-IR survey identified some young substellar objects at the periphery of the R CrA cluster (Kanai et al., in prep).

\begin{figure}[ht]
    \centering
    \includegraphics[height=8cm]{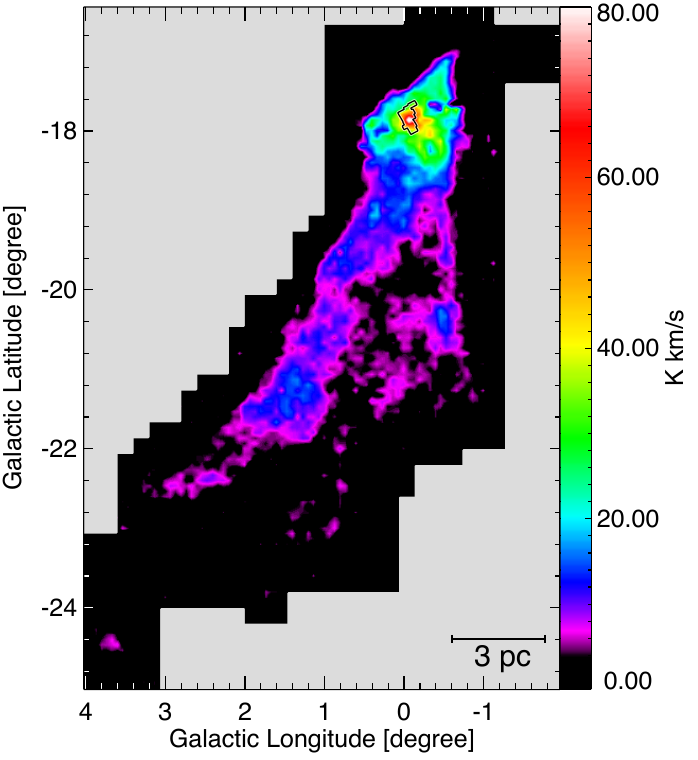}
    \includegraphics[height=8cm]{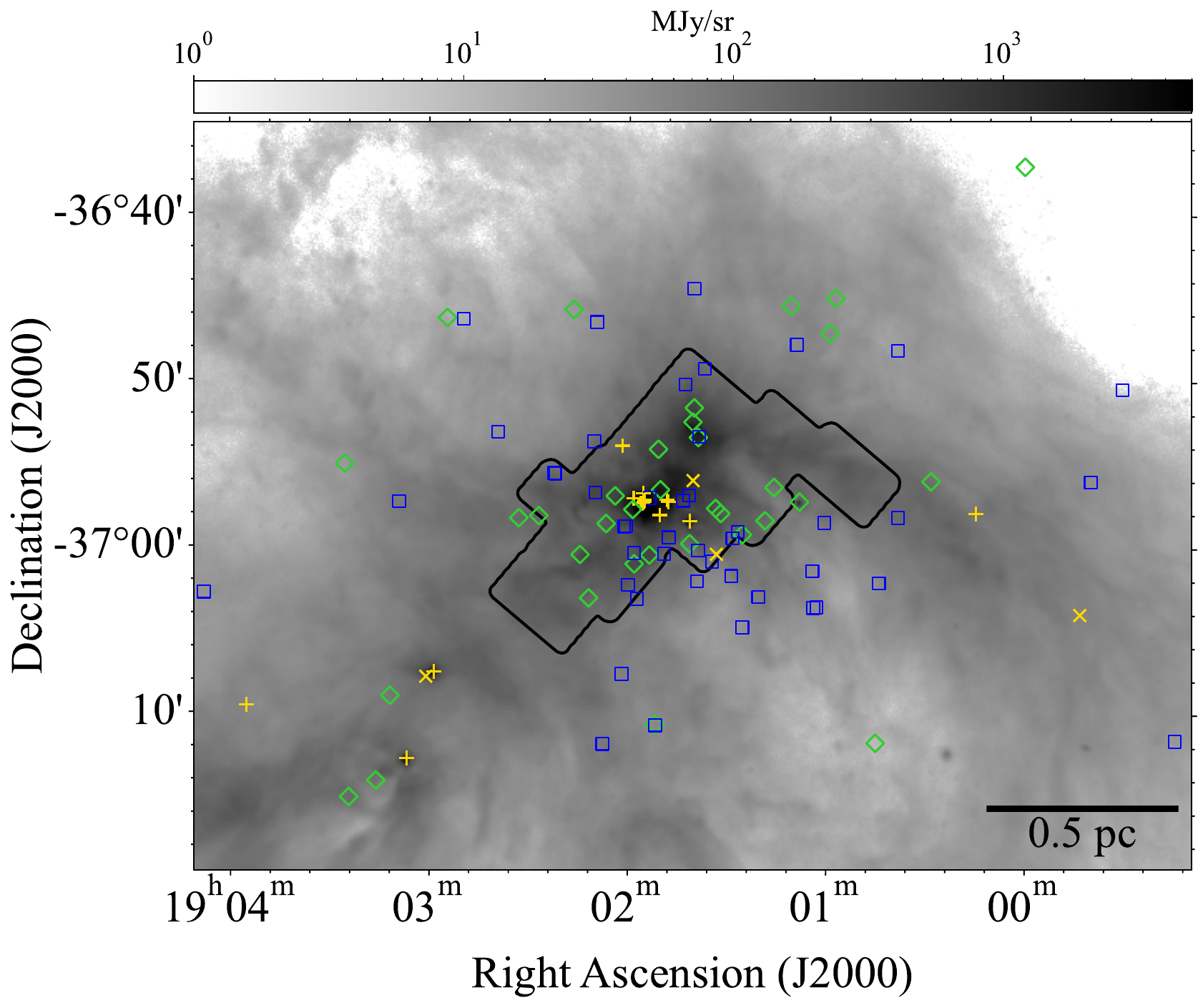}
    \caption{
    left: Integrated intensity map of the $^{12}$CO $J=$ 1--0 emission line covering the entire R CrA cloud by the NANTEN telescope (unpublished data), shown in the galactic coordinate system. right: Distribution of the dust continuum obtained with the Herschel telescope 250 $\mu$m, in the equatorial coordinates; the area surveyed by the ALMA ACA is indicated by the black line enclosure. The YSOs identified by \citet{peterson11} are indicated by symbols of yellow pluses for Class 0/I, yellow crosses for the flat spectrum, green diamonds for Class II, and blue squares for Class III objects.}
    \label{fig:large-scale_map}
\end{figure}

The morphology of the molecular gas has a characteristic head-tail/cometary structure of about more than 5 deg long corresponding to $> 13$ pc as shown by extinction \citep{rossano1978, cambresy1999}, suggesting an external disturbance. Fig.\ref{fig:large-scale_map} left shows integrated intensity map of $^{12}$CO $J=$1--0 line observed by the NANTEN telescope (unpublished data). From this map, the total mass of the entire molecular gas is estimated to be 7500 $M_{\sun}$. This is consistent with past observations of \ion{H}{1} and OH \citep{cappa91}. About 2000 $M_{\sun}$ molecular gas is concentrated in the head region of $3 \times 4$ pc$^2$ area derived from this $^{12}$CO $J=$ 1--0 data. Based on the NANTEN and SEST telescope observations of C$^{18}$O $J=$1--0, the average density of molecular gas in the most active star-forming dense core is $\sim 3 \times 10^4$--$10^5$ cm$^{-3}$ and the mass is estimated to be about 50--100 $M_{\sun}$ \citep{harju93, yonekura99}. \citet{bresnahan2018} observed the R CrA cloud with the Herschel space telescope, showing the distribution of interstellar dust, including prestellar cores and embedded protostars at a resolution of $36\arcsec$. The clustered Young Stellar Objects (YSOs) are embedded in the dense clump, that is the target of the present observation (Fig.\ \ref{fig:large-scale_map} right). The identified filamentary structure is stretched long along the tail, but shows a complex entanglement at the head. All of the high-mass prestellar cores tend to be distributed in the vicinity of the filaments. How these structures further fragment and provide the initial conditions for cluster formation is an interesting question. It is the most effective approach is to observe molecular gas in the nearest cluster forming region of the R CrA cloud at high resolution.

The Atacama Large Millimeter-submillimeter Array (ALMA) has high resolution and sensitivity, making it a powerful tool for studying the fragmentation of dense gas. Mosaic observations with the Atacama Compact Array (ACA; aka the Morita array) are particularly useful for studying the structure of nearby extended molecular clouds thanks to the relatively wide field of view of the 7m array and the ability to cover zero spacing with the Total Power (TP) array. Similar observations have been performed for high-density regions of the Taurus molecular cloud revealing the internal structures of many prestellar cores, allowing them to make statistical studies and discovery of the first core candidates \citep{tokuda19MC5,tokuda20,fujishiro20}. On the other hand, snapshot observations of the Chamaeleon molecular cloud using only the 12m array did not succeed in detecting the prestellar cores \citep{dunham16}. These indicate the usefulness of the wide-field mosaic observation by ACA.

In this paper, we report the first results of the ALMA ACA Band 6 mosaic survey observations of the R CrA molecular clouds, mainly those from C$^{18}$O $J=$2--1, SO $J_N = 6_5$--$5_4$ emission lines and the 1.3\,mm continuum observations. In Section \ref{section:observations}, we describe the ALMA ACA observational properties and data analysis including filament identification. Section \ref{section:results} summarizes the distribution of the continuum emission and C$^{18}$O emission lines and the physical quantities of the identified filaments. In Section \ref{section:discussions}, we discuss the characteristics of the filamentary structures identified and their origins, comparing them with numerical simulations. Section \ref{section:conclusions} summarizes these results. In the accompanying part 2 paper by Fukaya et al.\ (in preparation), we present higher resolution observations by the ALMA 12m array and discuss the fragmentations of the filamentary structures identified in the present paper and ongoing star formations therein.

%% file: 2_Observations.tex
\section{Observations and Analysis}
\label{section:observations}

\subsection{ALMA ACA observations}

Observations were carried out in Band 6 using the Atacama Compact Array (ACA) standalone mode of the ALMA telescope (project code 2018.A.00056.S). The observed area is shown in Fig.\ \ref{fig:large-scale_map}, and the entire area of 175 arcmin$^2$ was divided into 12 rectangular fields, covered by mosaics of 1501 pointings with the 7m array.

In order to recover the missing flux of the extended emission, the total power (TP) array was also used. A emission-free point at ($\alpha_{\rm J2000},\ \delta_{\rm J2000}) = (18^{\rm h}58^{\rm m}38\fs3,\ -37\arcdeg39\arcmin22\farcs2$) was chosen as the off-position. The TP-array observations covered not the entire 7m-array target fields but partial fields of 129 arcmin$^2$, and the achieved sensitivity is different for each of them, because the executions of the project timed out in the cycle. Only 4 fields of the TP array observations were complete and 5 fields incomplete, while the rest of northern 3 fields were not observed at all.

The Band 6 receiver was tuned to detect $^{12}$CO, $^{13}$CO, C$^{18}$O ($J=$2--1) emission lines, N$_2$D$^+$ ($J=$3--2) emission lines, and the continuous emission. SO $J_N = 6_5$--$5_4$ emission line falls in the Spectral Window (SpW) for the C$^{18}$O lines, and was obtained simultaneously.
The velocity resolution of the spectra was 0.1 km s$^{-1}$ for $^{12}$CO and N$_2$D$^+$ and 1.6 km s$^{-1}$ for the other lines, and a bandwidth of 4 MHz was assigned to the continuum (two SpWs were used, both with a bandwidth of 2 GHz). 
Note that in this paper we only present the results of C$^{18}$O and SO line data and continuum to focus on the internal structures of the molecular cloud. The properties of lower density gas and cold and dense cores observed by the other lines will be reported elsewhere.

\subsection{Data reduction}

Data was reduced using CASA \citep{casa22} version 6.4.4-31. 
Standard calibrations provided by the ALMA observatory were applied to each measurement set in a total of 1501 pointings composed of 12 fields with the 7 m array and 9 with the TP array, and then we concatenated all of the visibility files and imaged them together. The imaging strategy is the same as the previously reported ACA large-scale surveys \citep{Tokuda21,Muraoka23}. 
We used \verb"tclean" task with the \verb"multi-scale" deconvolver \citep{Kepley20} to recover the extended emission.
We applied the natural weighting, and the CLEAN mask was automatically created using the \verb"auto-multithresh" procedure. The obtained images were exported to maps with $1\arcsec \times 1\arcsec$ grid spacings, while the synthesized beam has an extent of $7.89\arcsec \times 4.88\arcsec$ with PA = $-77.4\arcdeg$. 
After the imaging task for the 7m array alone data, a sensitivity of 0.67 mJy\,beam$^{-1}$ was achieved for the continuum and $T_{\rm rms} = 0.025$ K for the emission lines with a velocity resolution of 1.6 km s$^{-1}$ and 0.076 K for the SpW with 0.1 km s$^{-1}$ resolution. 
The TP-array sensitivity ranges from 1.1 mK to 6.2 mK. The TP array data for C$^{18}$O and SO were regridded at the same pixel scale as the 7m-array image and combined by the \verb"feather" task.

%% file: 3_Results.tex
\section{Results}
\label{section:results}

\subsection{Overall properties of the cloud}

\begin{figure}[ht]
    \centering
    \includegraphics[height=11cm]{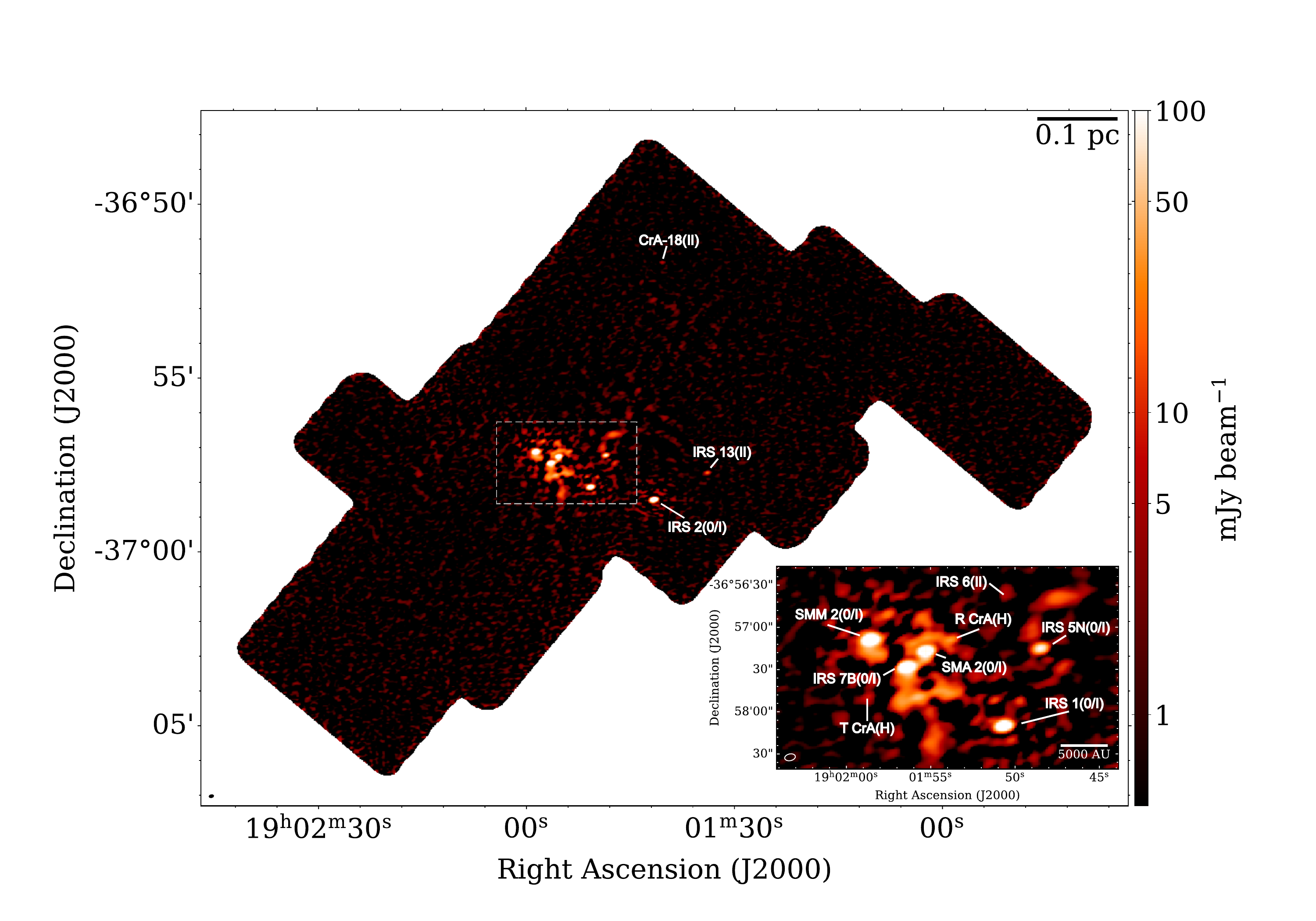}
    \caption{
    Intensity distribution of 1.3 mm continuum by the 7m array shown with logarithmic color scale. The inset in the lower right is an enlarged view of the white square in the center. For point sources, the names of the corresponding counterpart YSOs and their Classifications are denoted as 0/I and II in parentheses indicate Class 0/I and II, respectively, and H indicates Herbig Ae/Be star.
    }
    \label{fig:cont_map}
\end{figure}

\begin{figure}[ht]
    \centering
    \includegraphics[height=6.2cm]{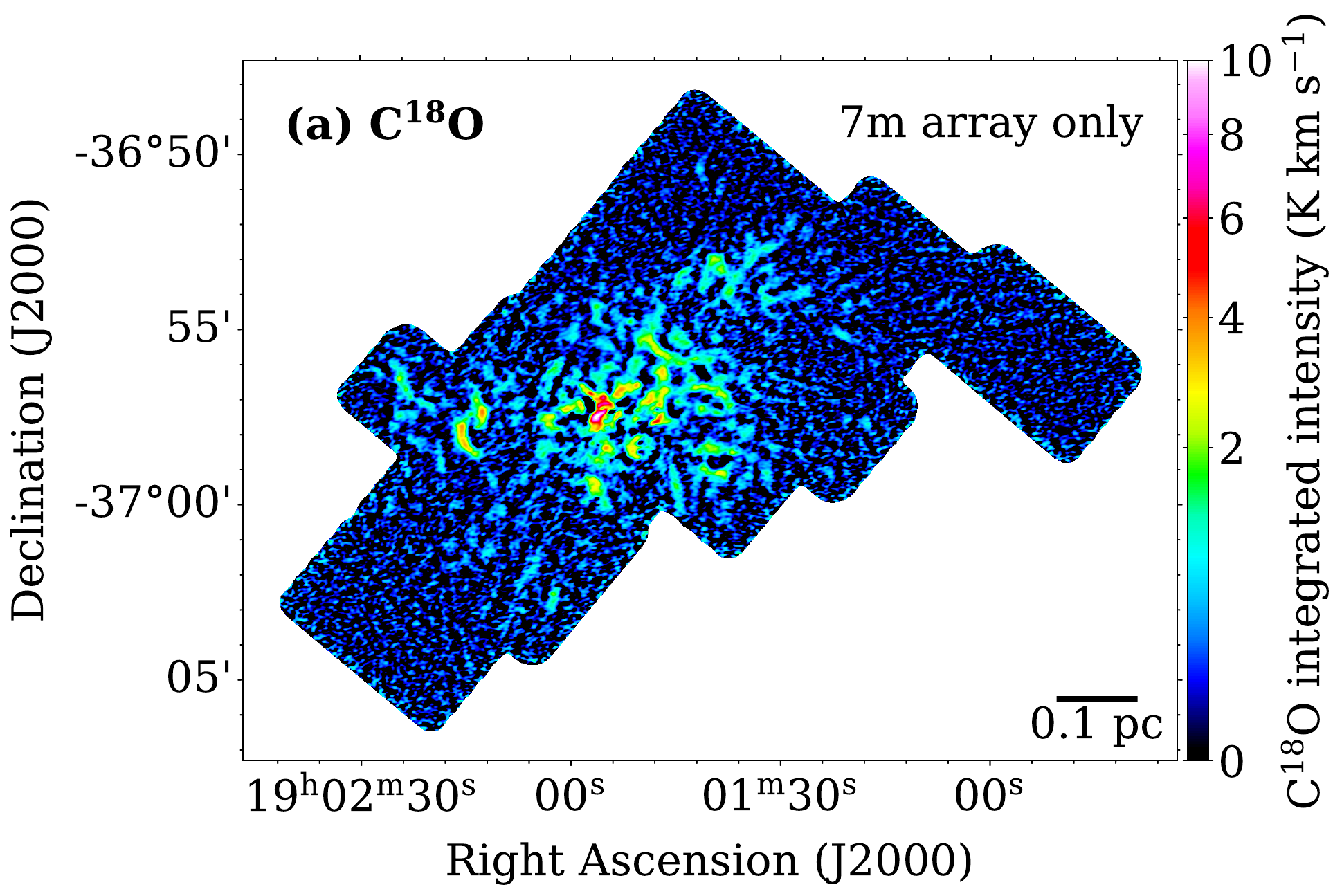}
    \includegraphics[height=6.2cm]{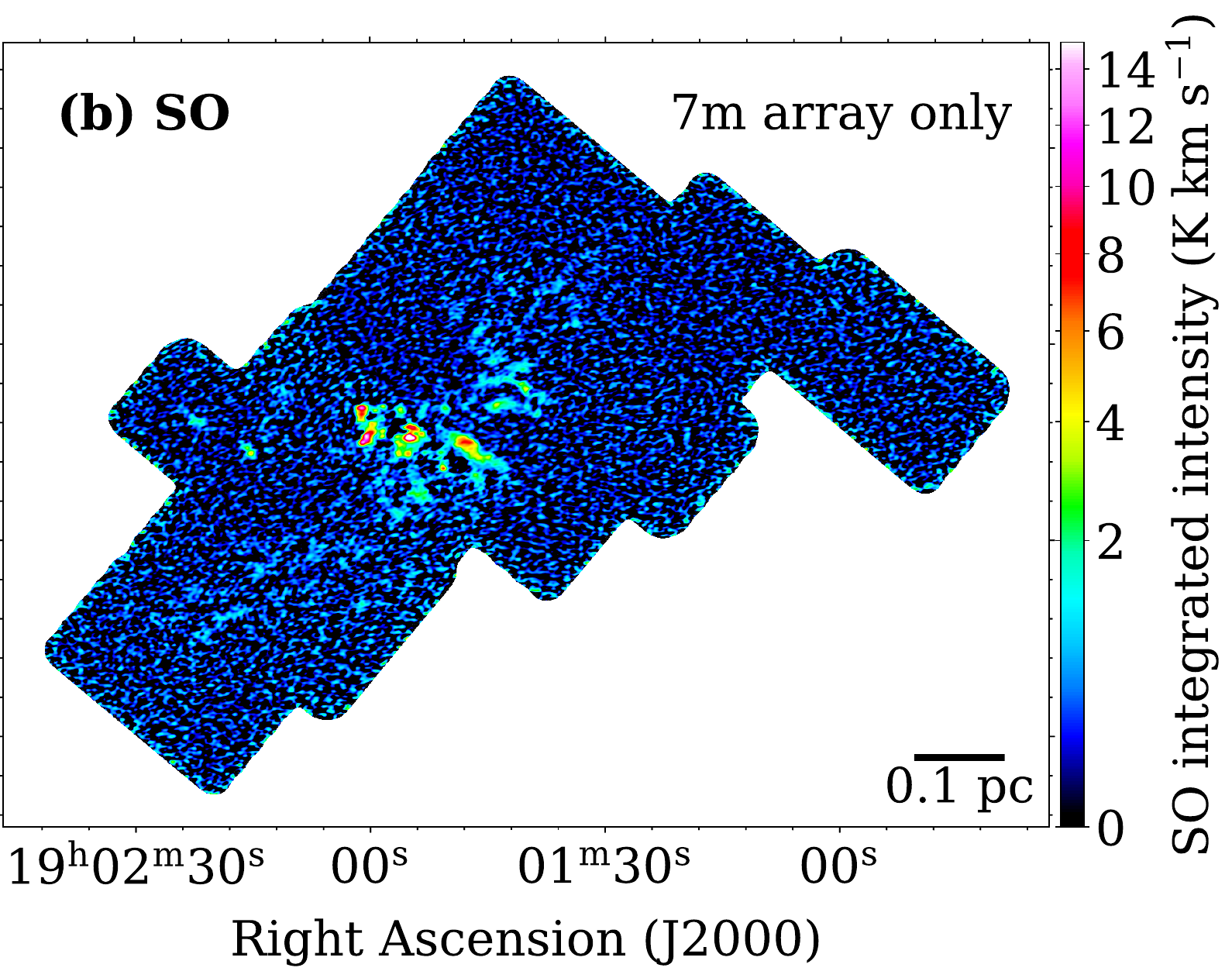}
    \includegraphics[height=6.2cm]{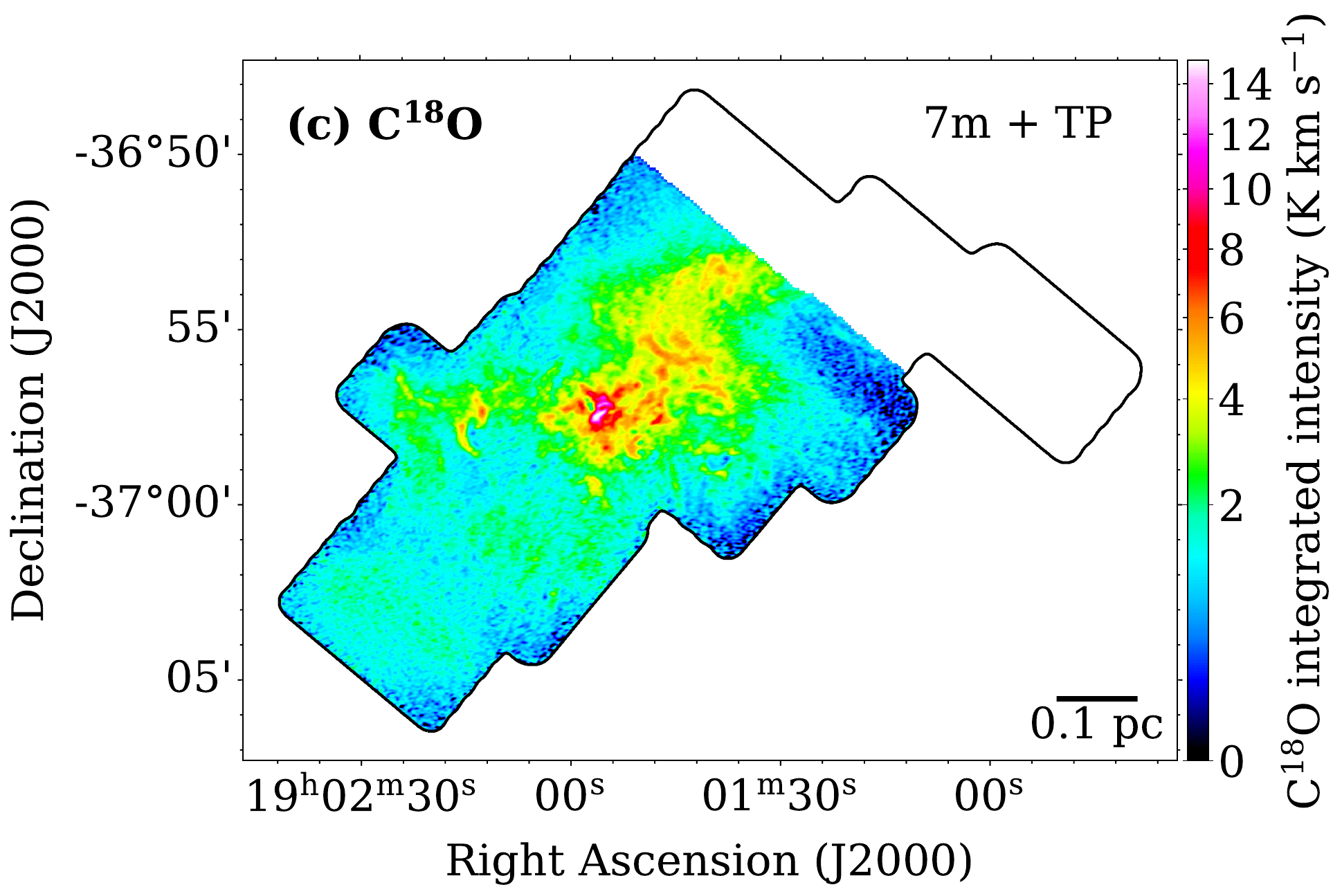}
    \includegraphics[height=6.2cm]{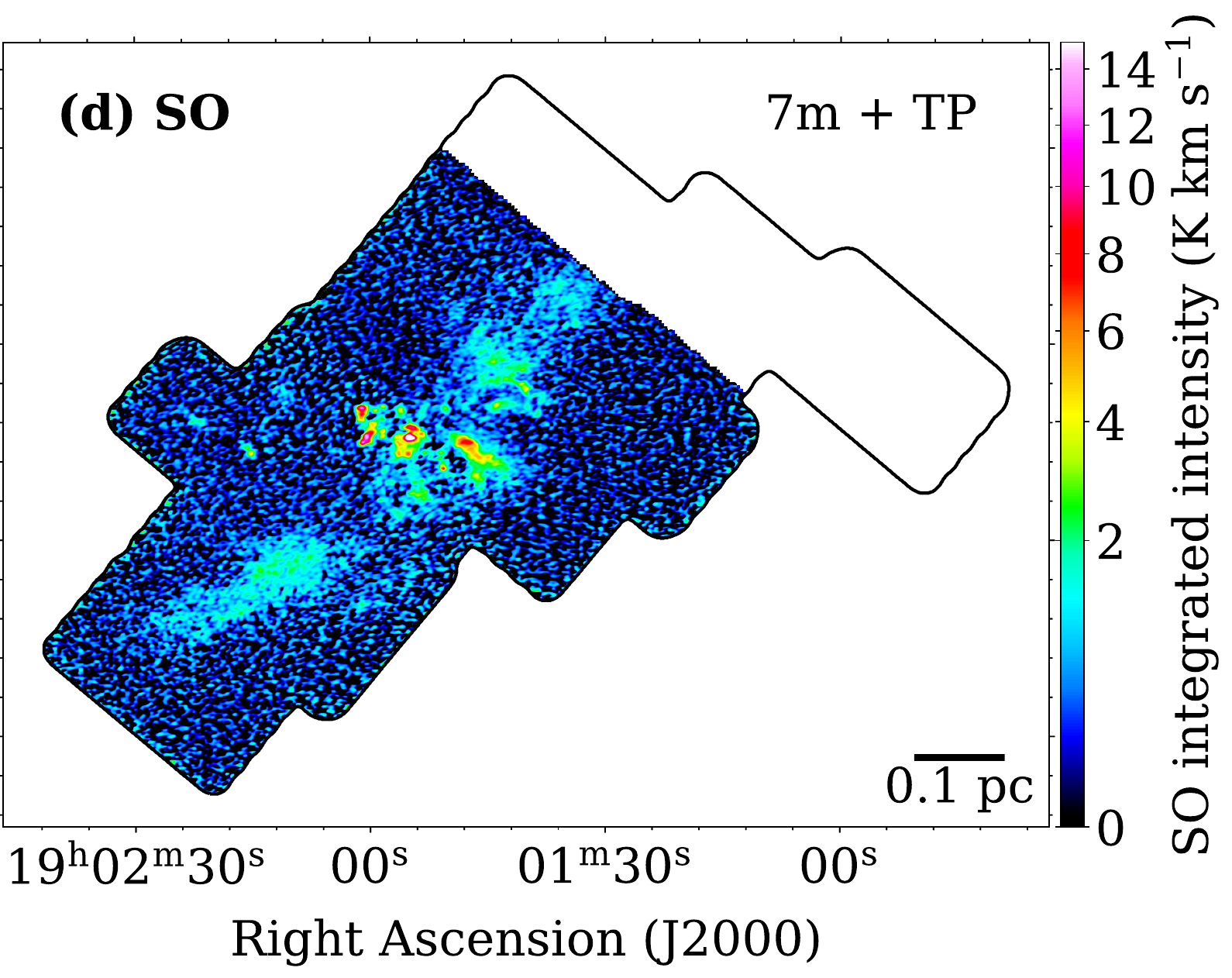}
    \caption{
    (a, b) Integrated intensity (moment 0) diagrams for C$^{18}$O, SO by the 7m array only data. (c, d) Same as (a, b) but with the TP array data combined. All color-coded intensities are displayed on a square root scale.
    }
    \label{fig:mom0_map}
\end{figure}

\begin{figure}[ht]
    \centering
    \includegraphics[height=6.2cm]{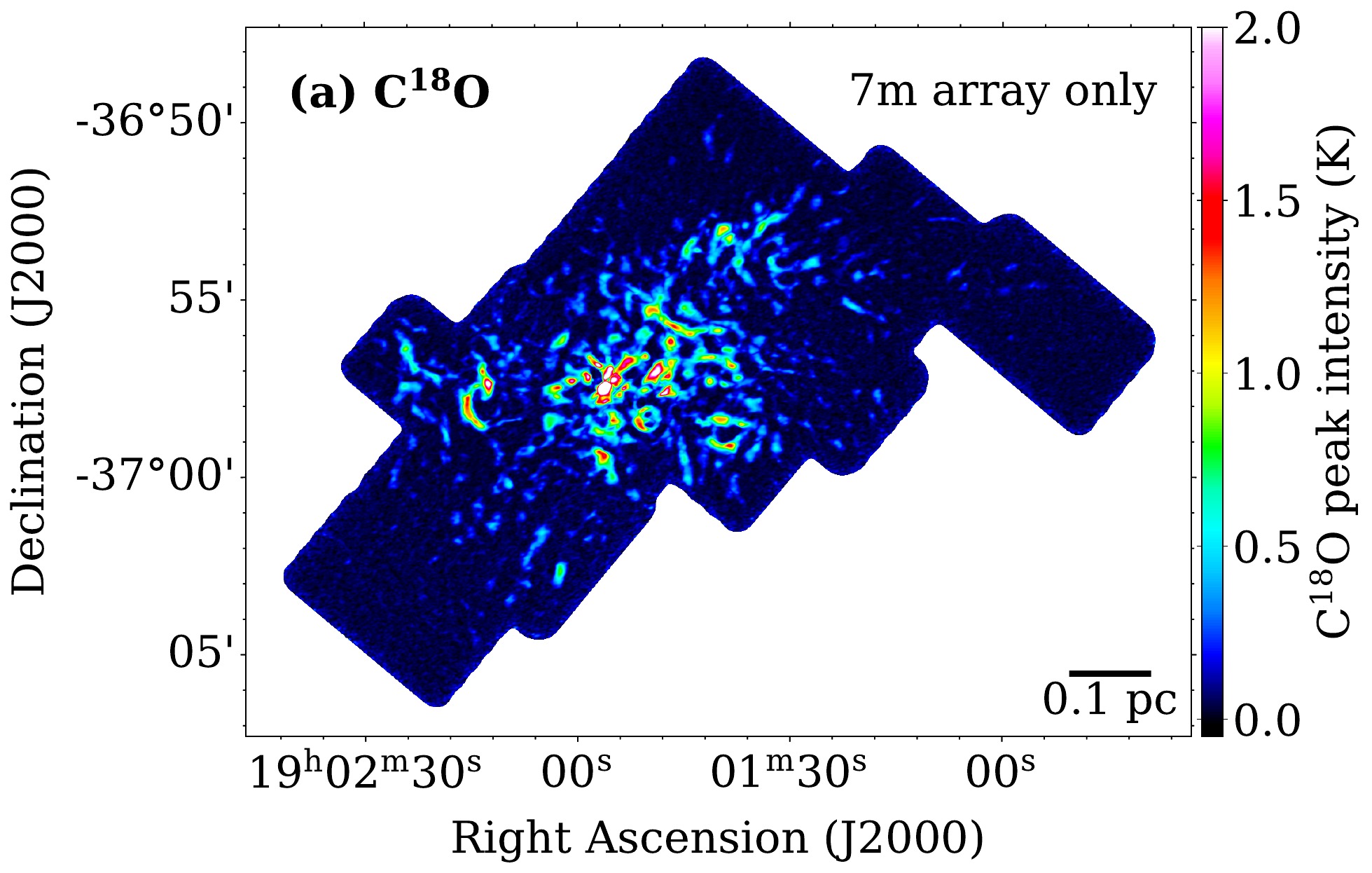}
    \includegraphics[height=6.2cm]{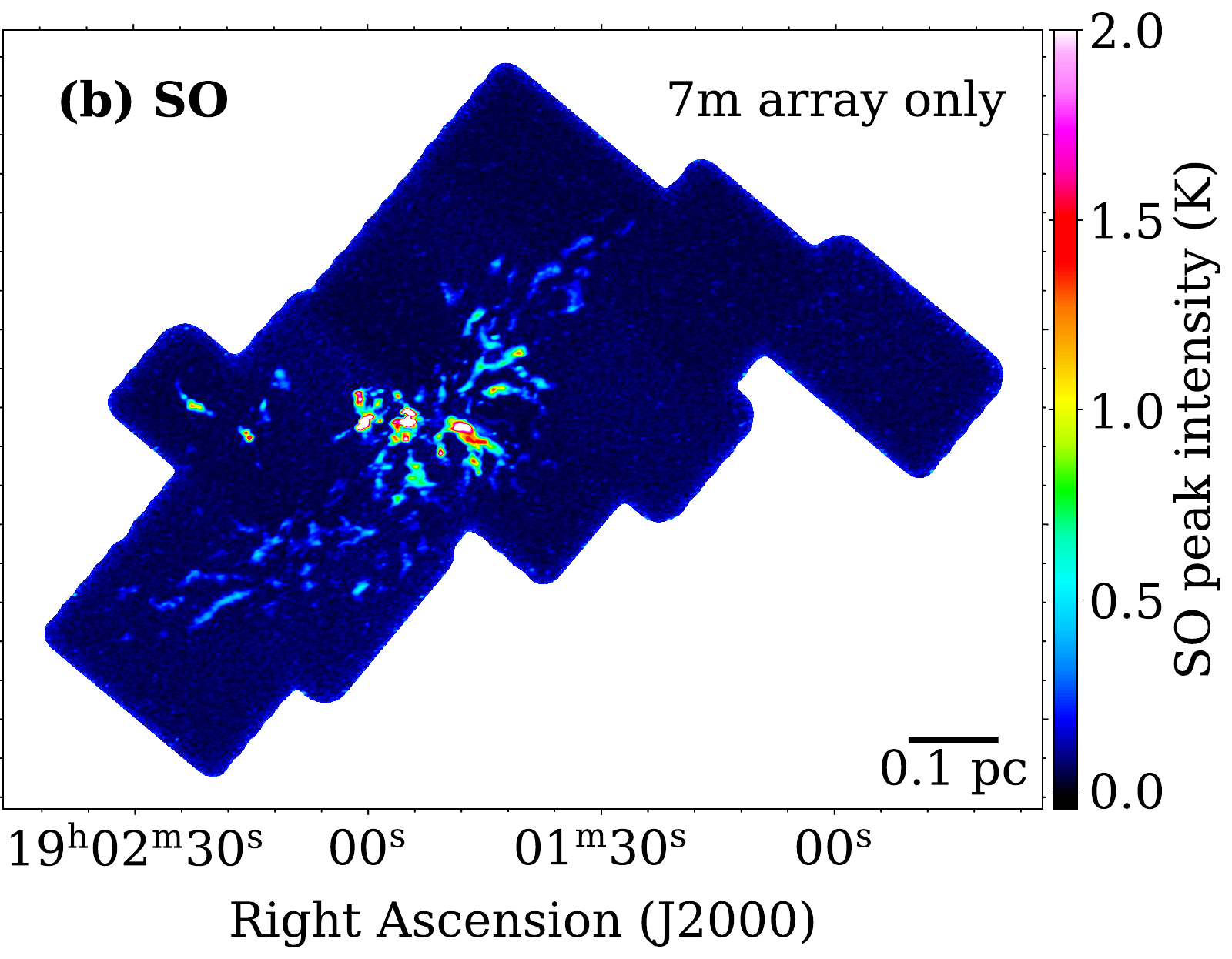}
    \includegraphics[height=6.2cm]{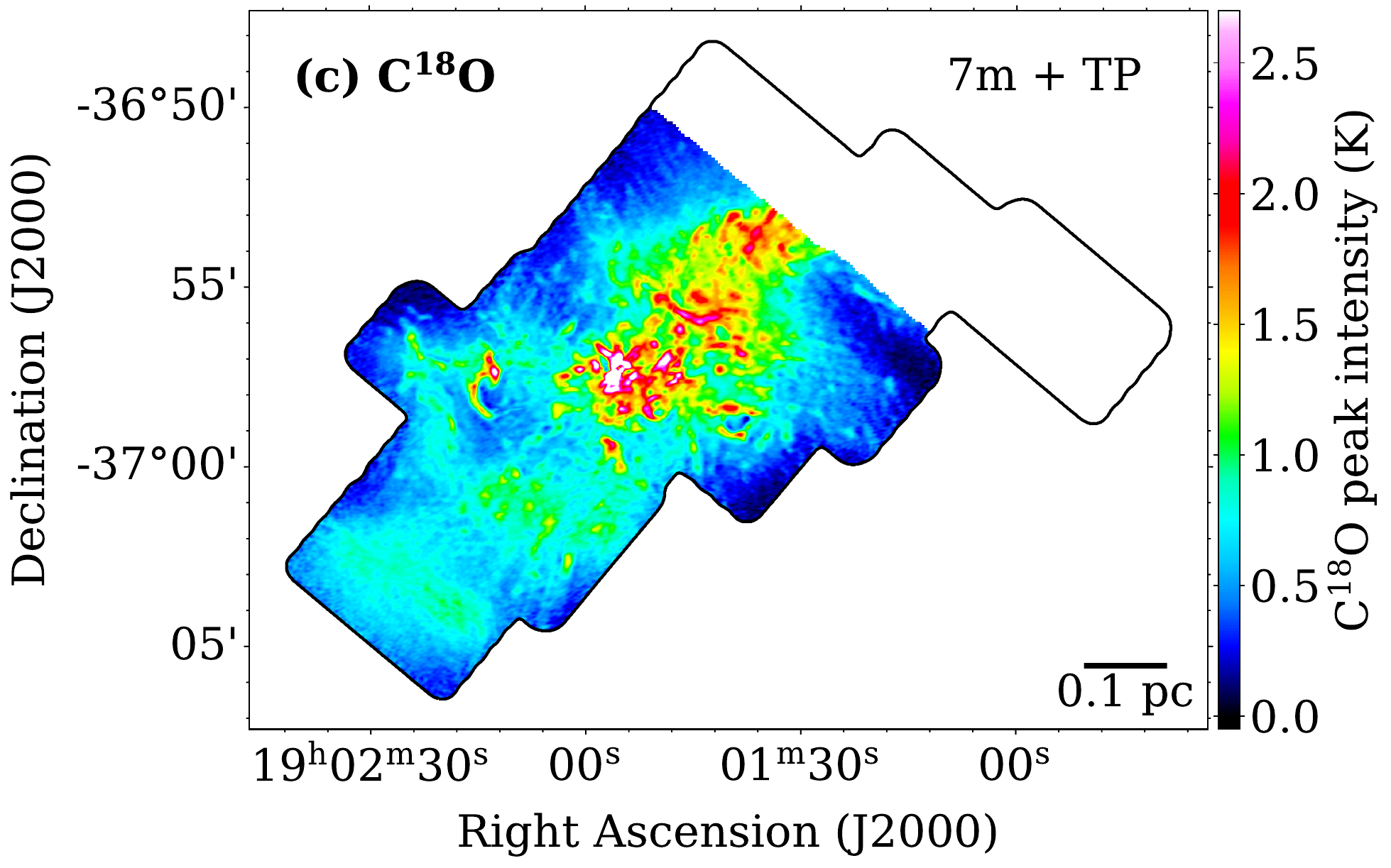}
    \includegraphics[height=6.2cm]{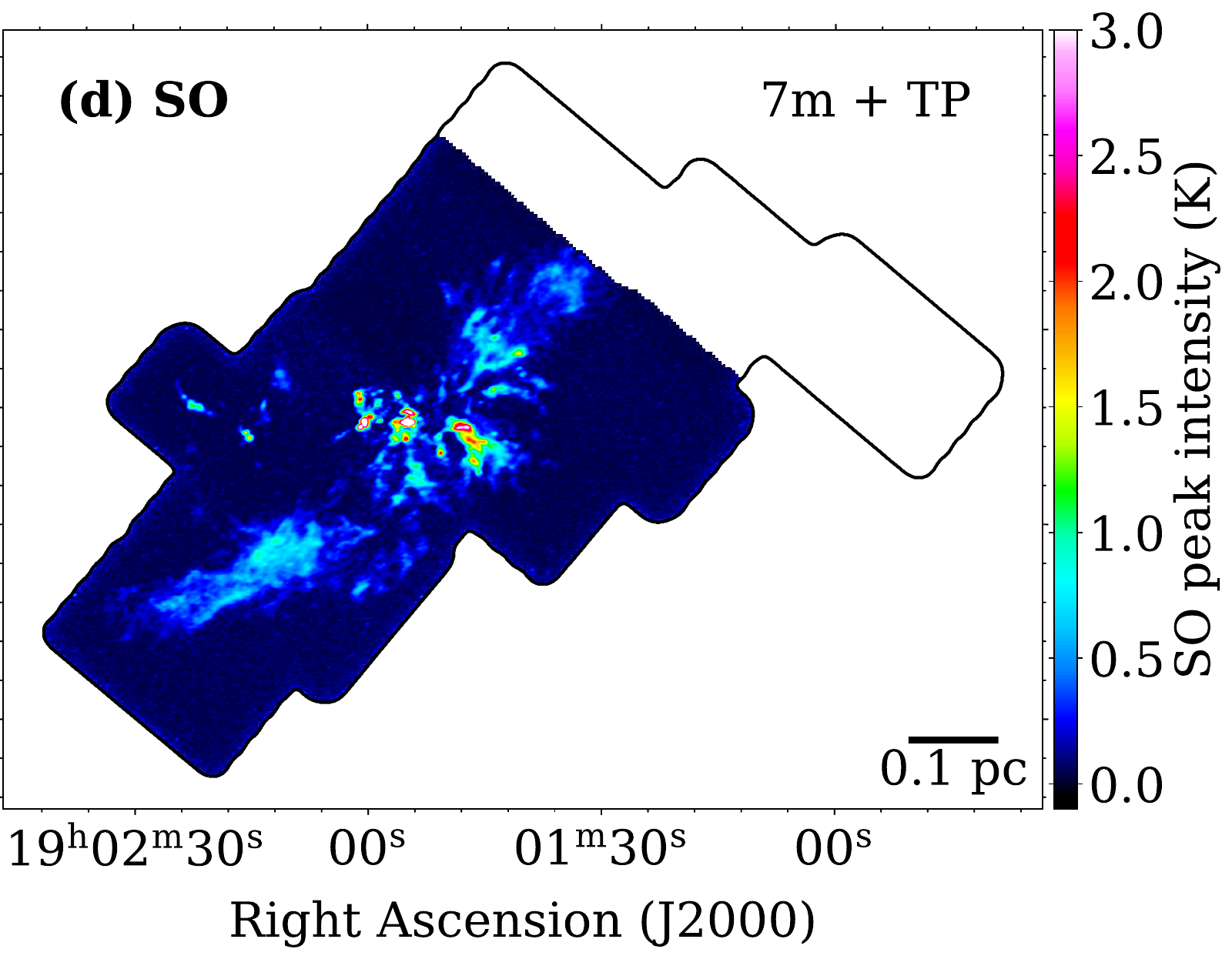}
    \caption{
    (a, b) Peak intensity maps for C$^{18}$O and SO by 7m array only data. (c, d) Same as (a, b), but combined with the TP array data combined. All color-coded intensities are displayed on a linear scale.
    }
    \label{fig:peakT_map}
\end{figure}

The continuum image obtained with the 7m array is shown in Fig.\ \ref{fig:cont_map}, while the integrated intensity maps obtained by combining the images of the 7m array and the TP array for C$^{18}$O and SO are shown in Figs.\ \ref{fig:mom0_map}a-d, and those of peak antenna temperature maps are shown in Figs.\ref{fig:peakT_map}a-d, respectively. 

Among the 11 Class 0/I sources identified by the Spitzer and SMA observations \citep{peterson11, groppi07}, 6 sources are clearly detected in the 1.3 mm continuum as the counterparts with the separations less than 2 arcsec between the 1.3 mm peak positions and the catalogued IR coordinates. In addition, 3 Class II sources and Herbig Ae/Be stars of T CrA and R CrA are detected, while Class III sources are not.
On the other hand, some continuum structures clearly extended more than the point spread function (PSF) exist especially near the center of the cluster. Some prestellar cores are likely to be detected, but the detection of filamentary structures connecting them is unclear. This is mainly due to resolution limitations, and it is difficult to judge whether they are superpositions of multiple sources or continuously connected filaments from the ACA results alone.
For the most active star forming region, IRS-7, higher resolution observations with the ALMA 12m array show clearly extended continuum emission in addition to the embedded point sources and prestellar cores. Details will be discussed in the part 2 paper.

The overall distributions of the C$^{18}$O and SO lines are quite different as seen in Fig.\ \ref{fig:mom0_map} and \ref{fig:peakT_map}. Both lines are intense toward the CrA cluster, but C$^{18}$O emission is more extended toward north-east with diffuse component surrounding the elongated filamentary structures. SO emission has, on the other hand, more clumpy and filamentary distributions in the middle and the north-west parts of the cloud. The diffuse and filamentary structure in SO is notable in the south-east part of the cloud where no C$^{18}$O emission is detected. 
In both the C$^{18}$O and SO lines, many filamentary structures are remarkable particularly for the 7m array only images as the diffuse extended components are filtered out. The ridges of the filamentary structures are, however, located on top of the extended structure in the combined images. These filamentary structures are especially abundant near the CrA cluster. Although some of the filamentary structures overlap each other in C$^{18}$O and SO, most of them appear to be independent structures.

Filamentary structures have been discovered in various molecular clouds ubiquitously with or without star formation by the Herschel Space Telescope. Interestingly, many of them have uniform widths of about 0.1 pc, and their origin has been of interest as a great mystery. In the case of the CrA clouds, most of the filaments identified by Herschel are distributed along the long elongation of the CrA cloud, while many filaments are entangled in the head section \citep{bresnahan2018}. The large-scale structures detected by the TP array are consistent with the distribution of the Herschel filaments, but those detected by the 7m array have much smaller-scale and regarded as the sub-structures within the filaments.
As an effect of the spatial filtering of the interferometric observations, it is possible that the sub-structure appears only in the 7m array as an artifact. For example, if an extended cloud has a structure with sharp edges, they could be detected as apparent structure. This can be confirmed by comparing the combined image with the TP array image. Comparisons of Fig.\ \ref{fig:peakT_map}a and \ref{fig:peakT_map}b, and \ref{fig:peakT_map}c and \ref{fig:peakT_map}d show that all the filamentary sub-structures found in the 7m array are embedded inside of the extended structures, rather than located at the edges (for more detail visual inspection, see Fig.\ \ref{fig:mom0rotate} in the Appendix \ref{sec:zoom-up}). This means that the extended emissions have smooth boundaries and there are no sharp edges. In other words, these sub-structures are not false structures, but can be interpreted as actually existing enhancements of column densities. For more details, see also the profile of spatial cuts in the column density (Fig.\ \ref{fig:column_cut}), which will be discussed quantitatively later.

\subsection{Properties of the filamentary substructures}
\label{section:filament}

Filament identification has been performed to extract the physical properties of the above sub-structures. 
For the C$^{18}$O and SO images reported here in detail, an algorithm for filament identification, FilFinder, was used \citep{koch15} to extract its structural features. The procedure is as follows.
For the peak antenna temperature maps, we set $3\sigma = 0.075$ K as the detection limit and the minimum size of the structure as 600 pixels (1 pixel = $1\arcsec \times 1\arcsec$), and the areas below these threshold values were masked. 
FilFinder then traces the ridge of intensity and define ``skeletons" if the ridge has a length longer than 5 times the beam size. We then divided the identified skeletons at the junctions into pieces. After discarding short branches less than 3 times the beam size, we ended up with 101 C$^{18}$O and 37 SO branches defined. 
This means that the length of a skeleton ridge can be a few times larger than that of the branch if the skeleton has bifurcation points and break up into branches. As applying Gaussian fitting to the branches across the long axes, we obtained successful estimations of widths for 95 and 37 branches of C$^{18}$O and SO, respectively. 
It should be noted, however, that FilFinder provides an input parameter, \verb"adapt_threth" (0.006 pc in this case), which is an estimate of the filament width to be identified, so there remains a bias in the extracted filamentary structure width. In addition, the \verb"smooth_size" (0.003 pc in this case) were set to eliminate small noise fluctuations when creating the mask. 
The widths of thus identified filamentary structures were measured by the algorithm implemented in FilFinder. They were obtained by fitting the intensity distribution perpendicular to the ridge, assuming the Gaussian distributions.

Note that the velocity resolutions for the C$^{18}$O and SO emission lines are 1.6 km s$^{-1}$, and structures with small velocity dispersion were not detected. Also, structures with different radial velocities that may overlap on the line-of-sight are not separated. These issues are discussed in detail in Fukaya et al. (part 2 paper) with higher resolution 12m-array data.

We applied the FilFinder algorithm to the 7m array C$^{18}$O and SO peak $T$ maps, focusing on the data from the 7m array alone as the small-scale structures emerge with higher contrast on the images than on the TP alone and the combined images. Regarding the derivation of physical quantities, we analyzed the 7m array alone and the 7m+TP combined images based on the 7m array-based identified spines, as detailed in Section~\ref{Sec:NH2}. Technically, the use of the Peak $T$ maps, not the moment~0 maps, is advantageous as it allows us to exclude ambiguity related to velocity-range selection. The peak $T$ maps have better signal-to-noise ratio than the integrated intensity maps that incluse noise due to integration effects. While some velocity differences are observed between spatially adjacent filaments, the overlap of more than two velocity components in the same line-of-sight direction is not prominent, making the 2D map setting suitable for capturing the global trend in the observed region. 

\begin{figure}[ht]
    \centering
    \includegraphics[height=6.2cm]{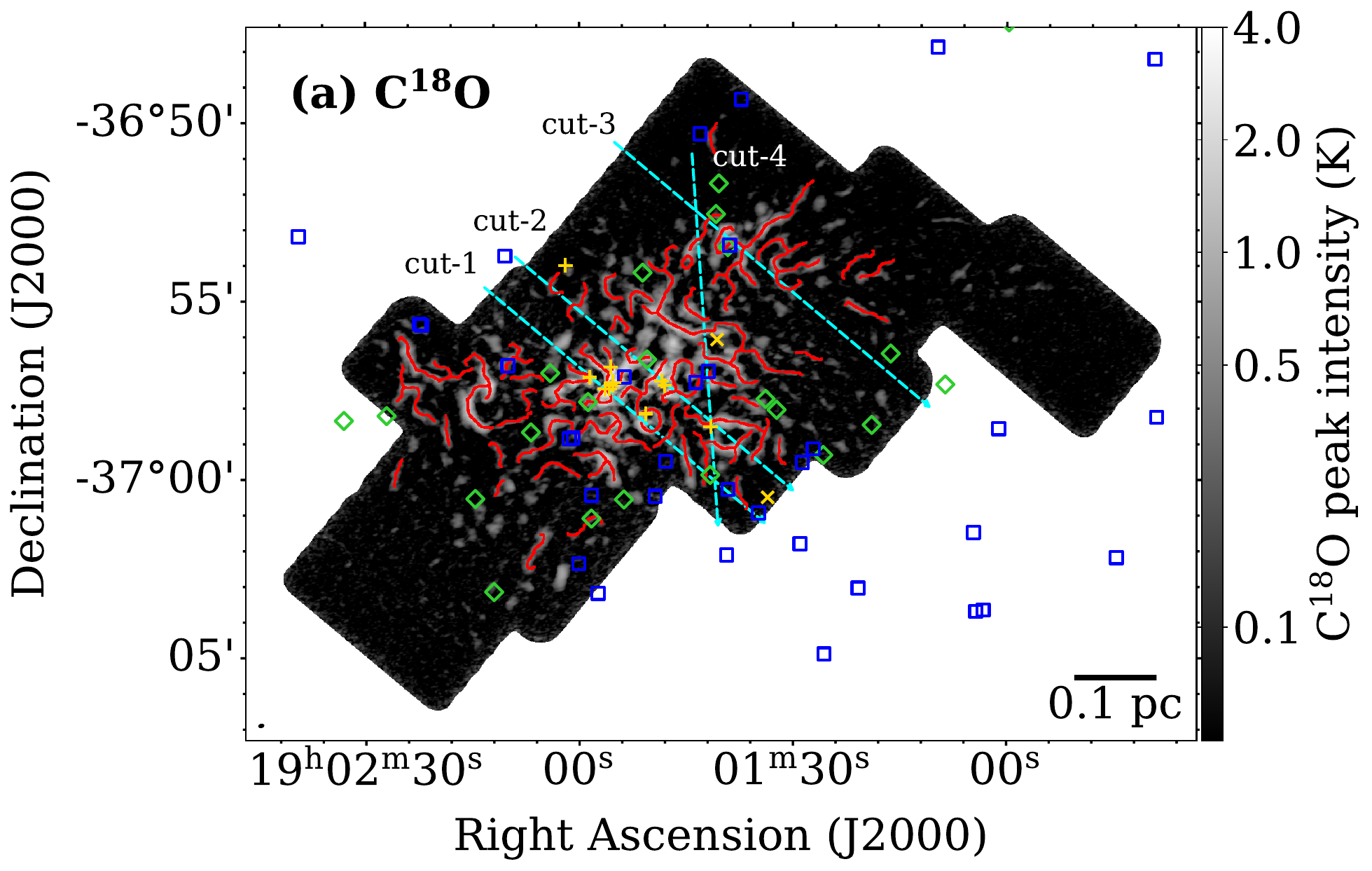}
    \includegraphics[height=6.2cm]{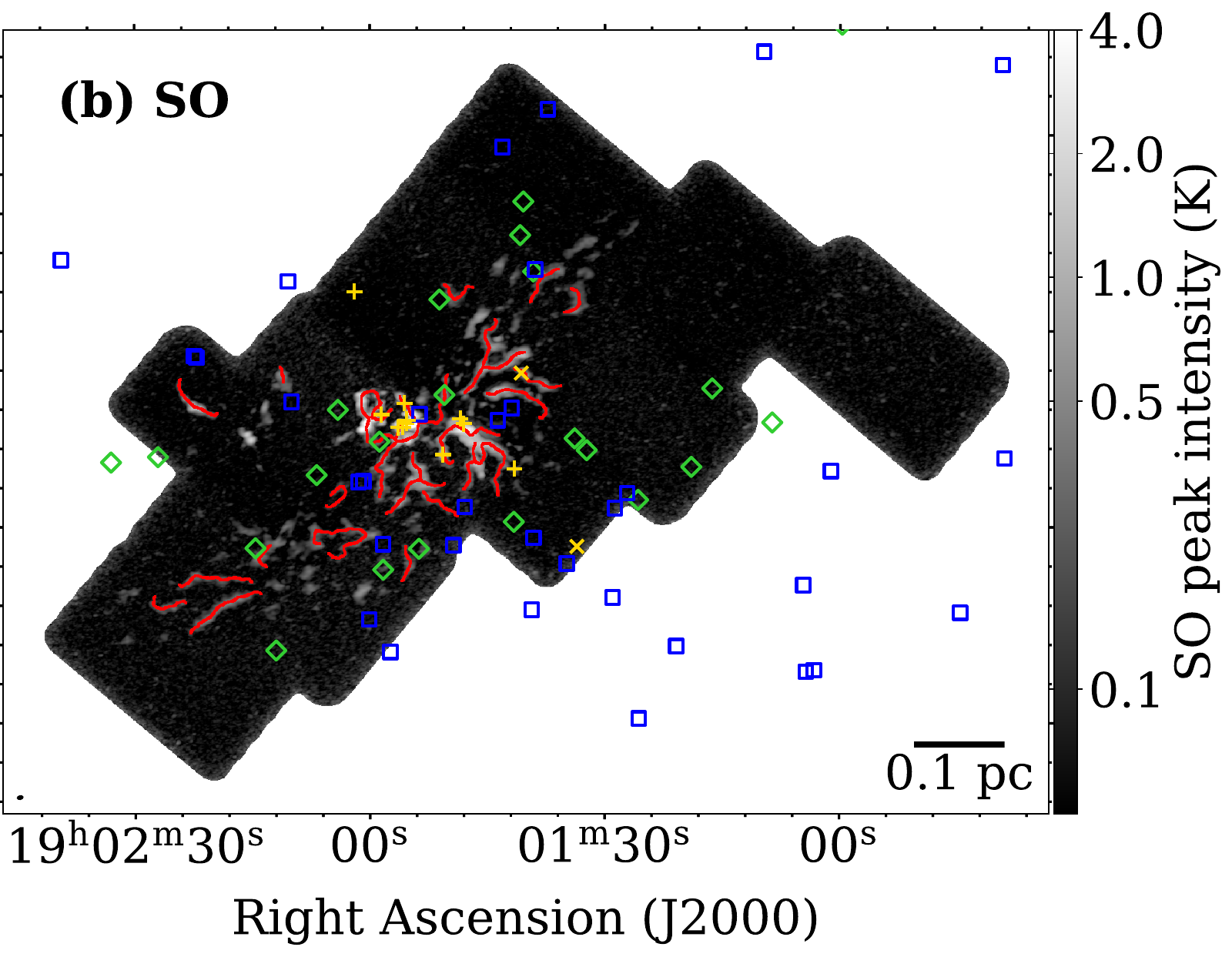}
    \caption{
    Filament identification results for (a) C$^{18}$O and (b) SO by FilFinder. The ridges of the identified filamentary structures are indicated by red lines on the background  C$^{18}$O peak $T$ map in grayscale. The cyan broken lines in (a) denote the 4 cut lines on which the column density distributions are shown in Fig.\ \ref{fig:column_cut}. The symbols are the same as in Fig.\ \ref{fig:large-scale_map} right.
    }
    \label{fig:finfinder}
\end{figure}

As a result, 101 C$^{18}$O and 37 SO structures were identified (Fig.\ \ref{fig:finfinder}). Both of them are especially abundant in high-density regions of the molecular clouds where star clusters are present, while the C$^{18}$O filamentary structures are also widely distributed in the surrounding regions. The directions of the filaments are almost randomly distributed in high-density regions. On the other hand in the low-density peripheral regions, they extend radially from the high-density center. The SO filaments, in particular, are distributed along the large-scale elongation of the cloud.

The C$^{18}$O and SO structures partially overlap. Therefore we checked them by eye and count them as the number of overlaps if the C$^{18}$O and SO structures have partially common ridges and do not cross each other. As a results, 19 out of the 110 C$^{18}$O structures overlap with the SO structures, while conversely 16 out of the 37 SO structures do. Because of the high critical density of the SO line, it is expected that the C$^{18}$O traces more structures with lower density, and those detected in SO are denser than the others. Indeed, many structures near the center of the molecular clump are detected in both lines, while those in the outskirts are only detected in C$^{18}$O. Exceptions are those located in the south-east part of the cloud, where diffuse and filamentary SO emission is detected. This is likely due to the effect of CO freeze-out to be discussed in Section\ \ref{subsection:feathers}.

Comparing these distributions with those of the YSOs, spatial associations have no clear trend. Majority of the Class 0/I protostars are located on the local peaks of C$^{18}$O with significant SO detections, while not all of such C$^{18}$O peaks are associated with protostars. 
Overall, only 11 and 6 out of the 101 C$^{18}$O and 37 SO filamentary structures, respectively, are associated with YSOs. On the other hand, a few YSOs including the Class 0/I protostar IRS 7B are concentrated at the peak position of C$^{18}$O on the filamentary structure near the center of the cluster. Note that the IRS 7 region in particular will be discussed in detail in the part 2 paper with higher resolution data by the 12m array.

\begin{figure}[ht]
    \centering
    \includegraphics[width=7cm]{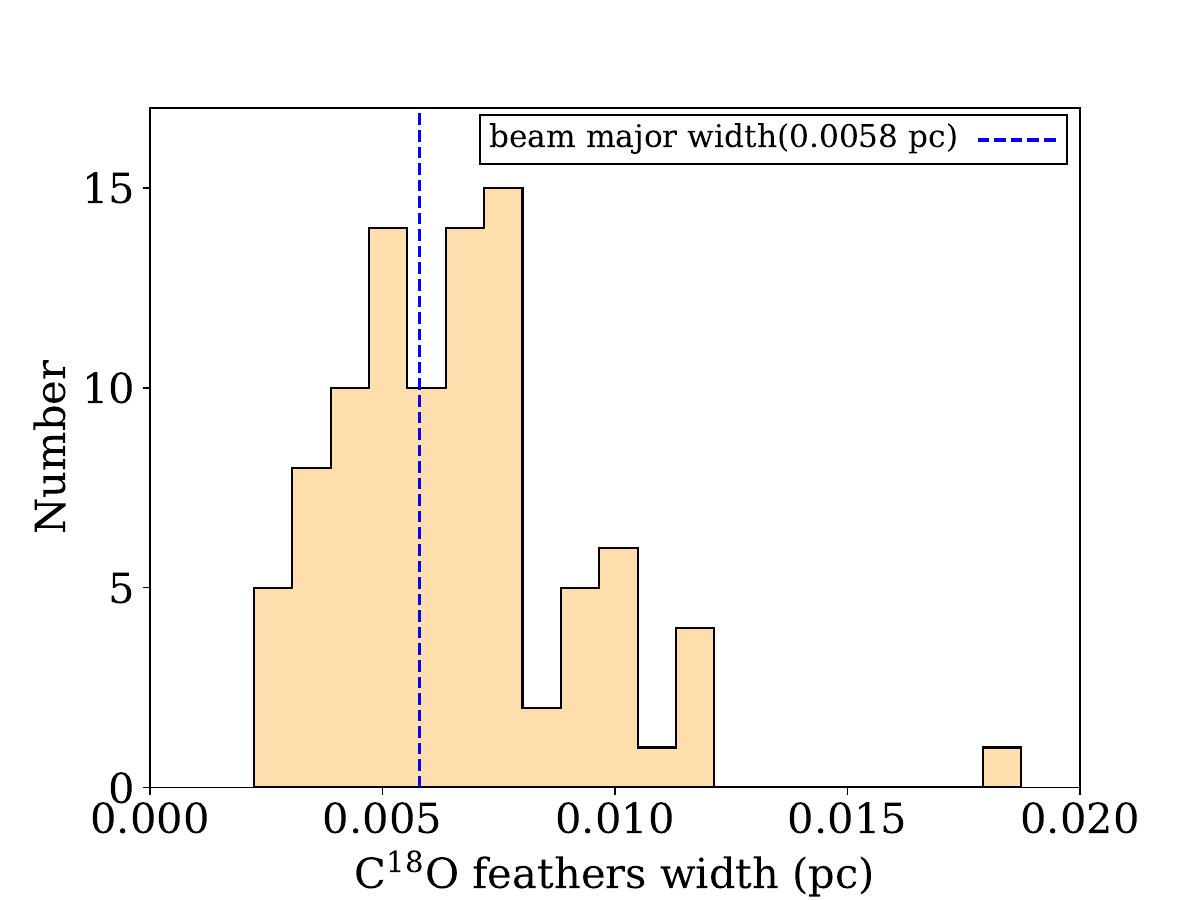}
    \includegraphics[width=7cm]{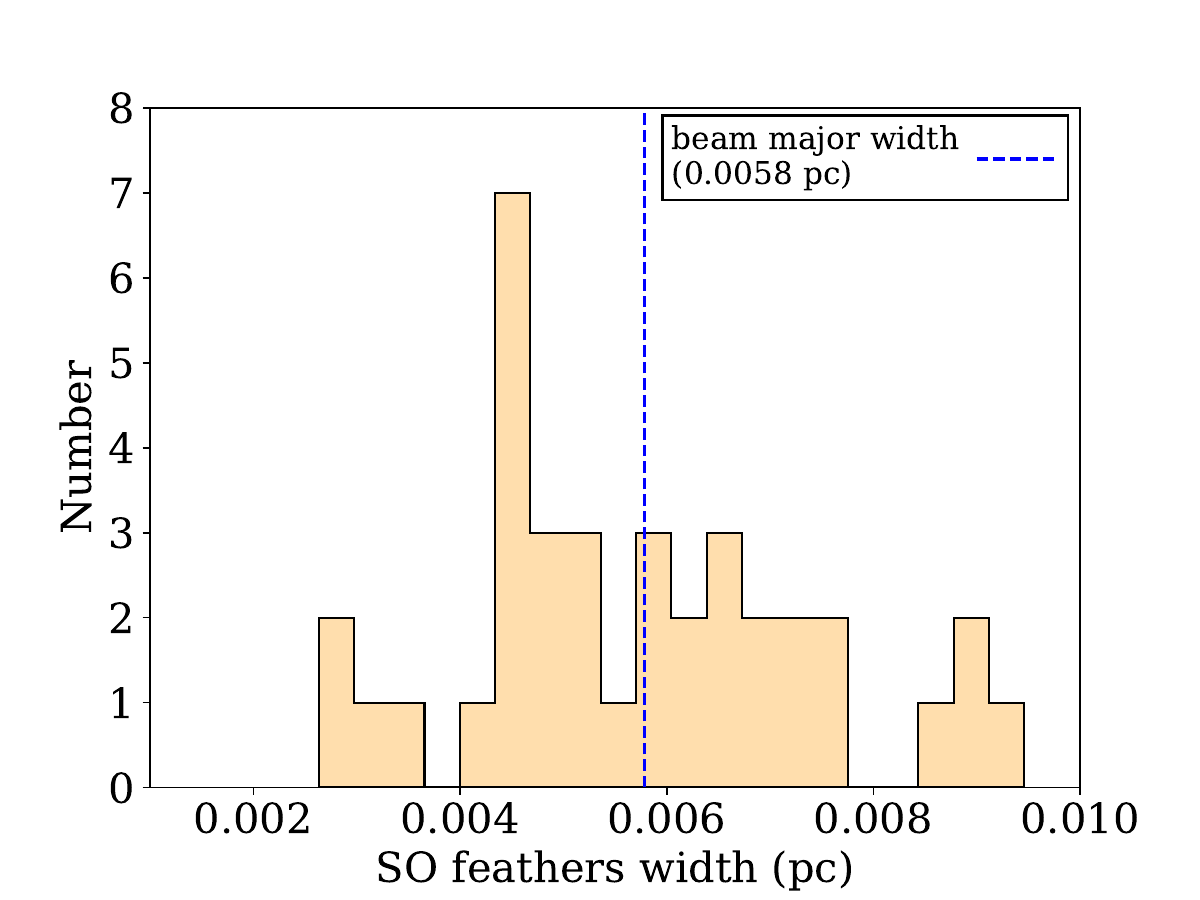}
    \includegraphics[width=7cm]{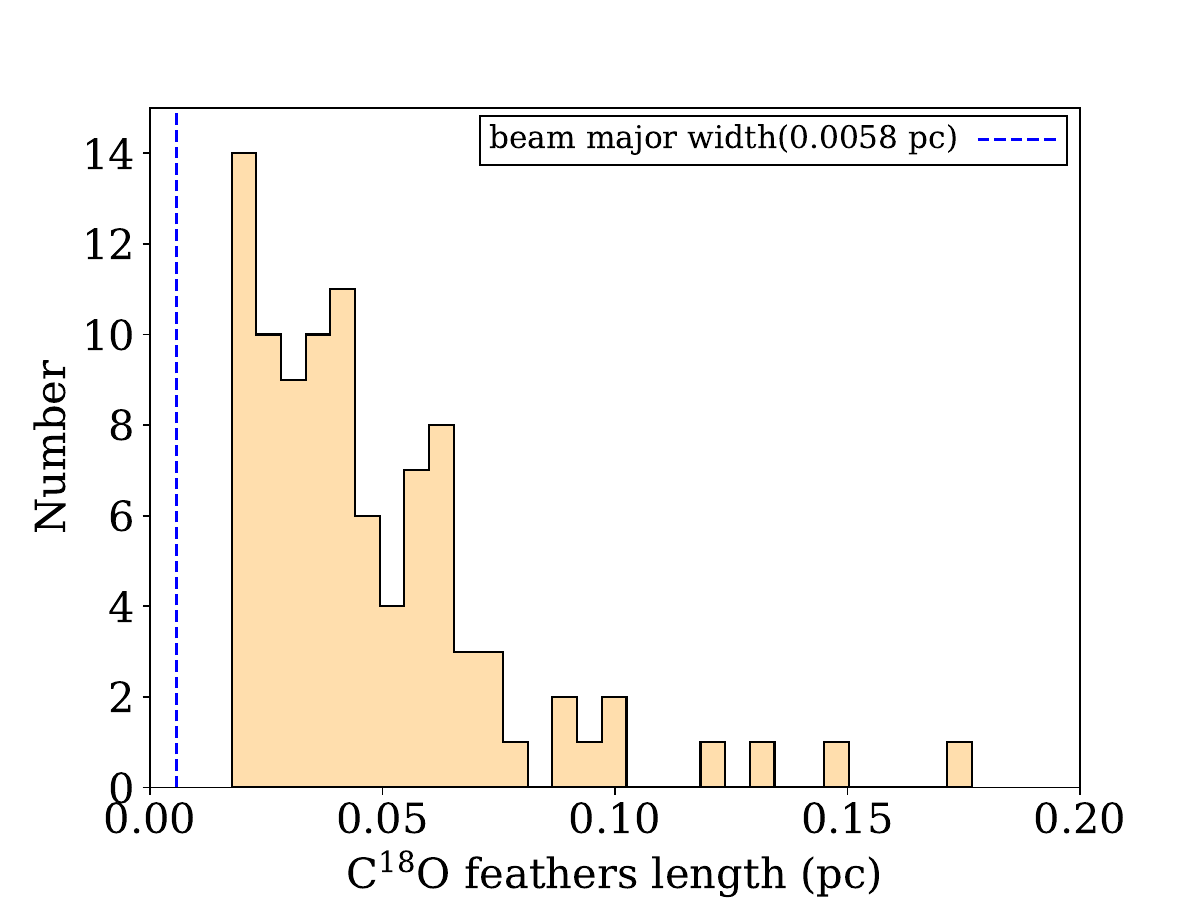}
    \includegraphics[width=7cm]{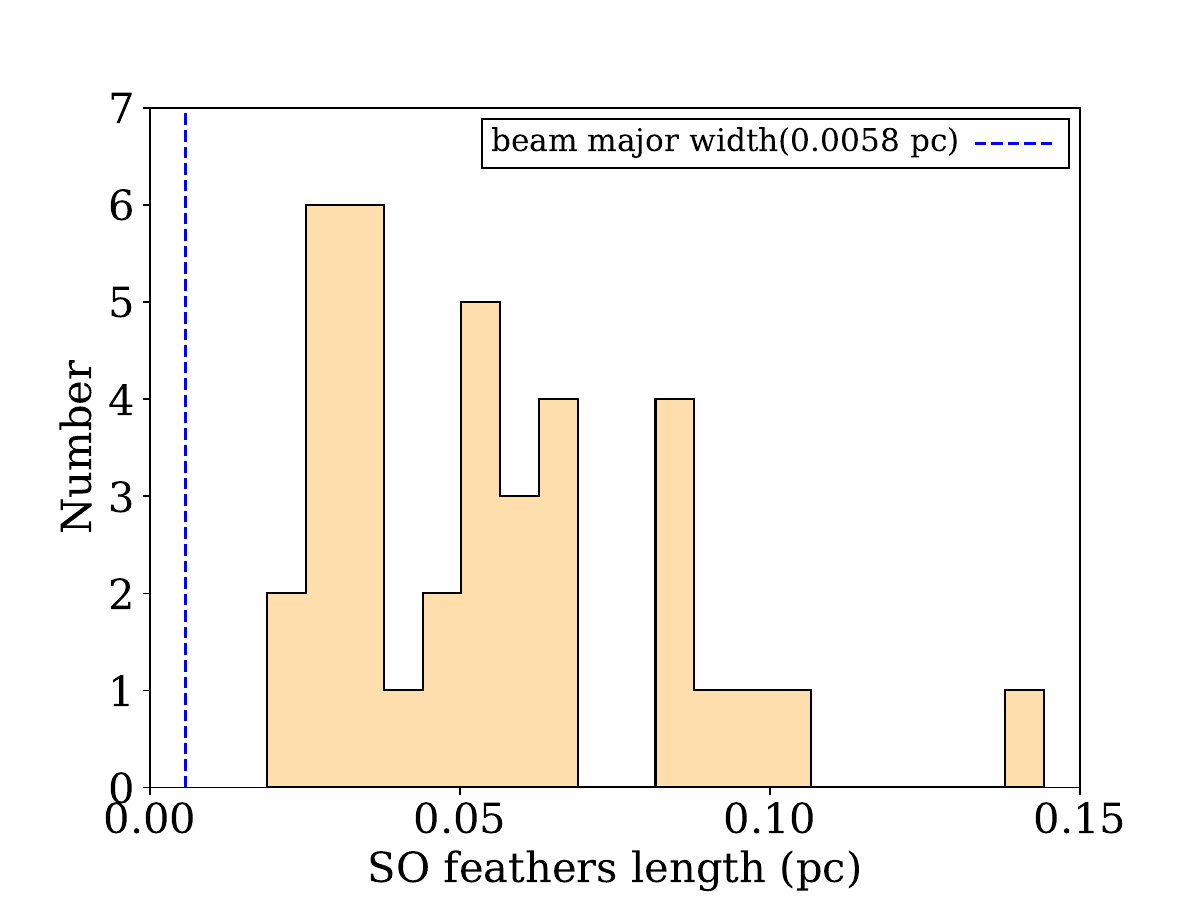}
    \caption{
    Histograms of the width (left) and length (right) of the filamentary structures identified in C$^{18}$O and SO. The blue vertical broken lines indicate the length of the major axis of the synthetic beam.
    }
    \label{fig:fil_width}
\end{figure}

The histograms of the widths and lengths of the C$^{18}$O and SO filamentary structures (branches) are shown in Fig.\ \ref{fig:fil_width}. Most of them have slightly larger widths than the beam size, with medians of 0.0064\,pc (1300 AU) and 0.0055\,pc (1100 AU), respectively, and thus they are only marginally resolved. Some of them are observed with higher resolution by the ALMA 12 m array and the widths are not significantly different from the ACA-only measurements, which will be reported in detail in the part 2 paper. The maximum width is 0.01 pc, and thus these structures are qualitatively different from the filaments with $\sim 0.1$ pc width defined by Herschel's observations.

\subsection{Column density distributions and embedded feathers}\label{Sec:NH2}
From the obtained C$^{18}$O intensity distributions (the 7m array only and the TP array combined), we obtained the molecular hydrogen column density $N({\rm H_2})$ distribution as follows. First, the C$^{18}$O emission lines were assumed to be in local thermodynamic equilibrium (LTE), and the excitation temperature ($T_{\rm ex}$) was assumed to be constant at 15 K, referring to the Herschel's dust temperature map and the peak $T_{\rm mb}$ of $^{12}$CO ($J=$1--0) line taken by NANTEN.
The column density of C$^{18}$O molecules $N({\rm C^{18}O})$ is obtained by the following equation. 
\begin{equation}
    N({\rm C^{18}O}) = 1.64 \times 10^{16} 
    \frac{\int \tau \, dV}{\exp\left (\frac{10.54}{T_{\rm ex}}-1 \right )} \ \  [{\rm cm^{-2}}]
\end{equation}
where $dV$ is the velocity channel width, $\tau$ is the optical depth at velocity $V$, which is obtained by the following equation.
\begin{equation}
    \tau = -\ln \left [1-\frac{T_{\rm mb}}{10.54}
    \left \{\frac{1}{\exp(10.54/T_{\rm ex})-1} 
    -0.02 \right \}^{-1} \right ]
\end{equation}
where $T_{\rm mb}$ is the brightness temperature of the C$^{18}$O line at velocity $V$ expressed in the unit of K. The calculations were performed for each velocity channel and integrated over the velocity to obtain $N({\rm C^{18}O})$ for each pixel.
Furthermore, assuming a constant molecular abundance of C$^{18}$O relative to H$_2$, $N ({\rm H_2})$ is derived as $N({\rm H_2})=[N({\rm C^{18}O})/(1.7 \times 10^{14})+1.3] \times 10^{21}$ [cm$^{-2}$] \citep{frerking82}.
It should be noted that the derived H$_2$ column density is subject to  underestimate due to the effect of CO freeze-out. CO molecules are supposed to be frozen onto the dust grain surface and depleted in cold and dense gas. As will be discussed in Section \ref{subsection:feathers}, CO freeze-out is likely to change the C$^{18}$O molecular abundance between the regions in the R CrA cloud. Thus $N({\rm H_2})$ derived here should be regarded as lower limit.

\begin{figure}[ht]
    \centering
    \includegraphics[width=8cm]{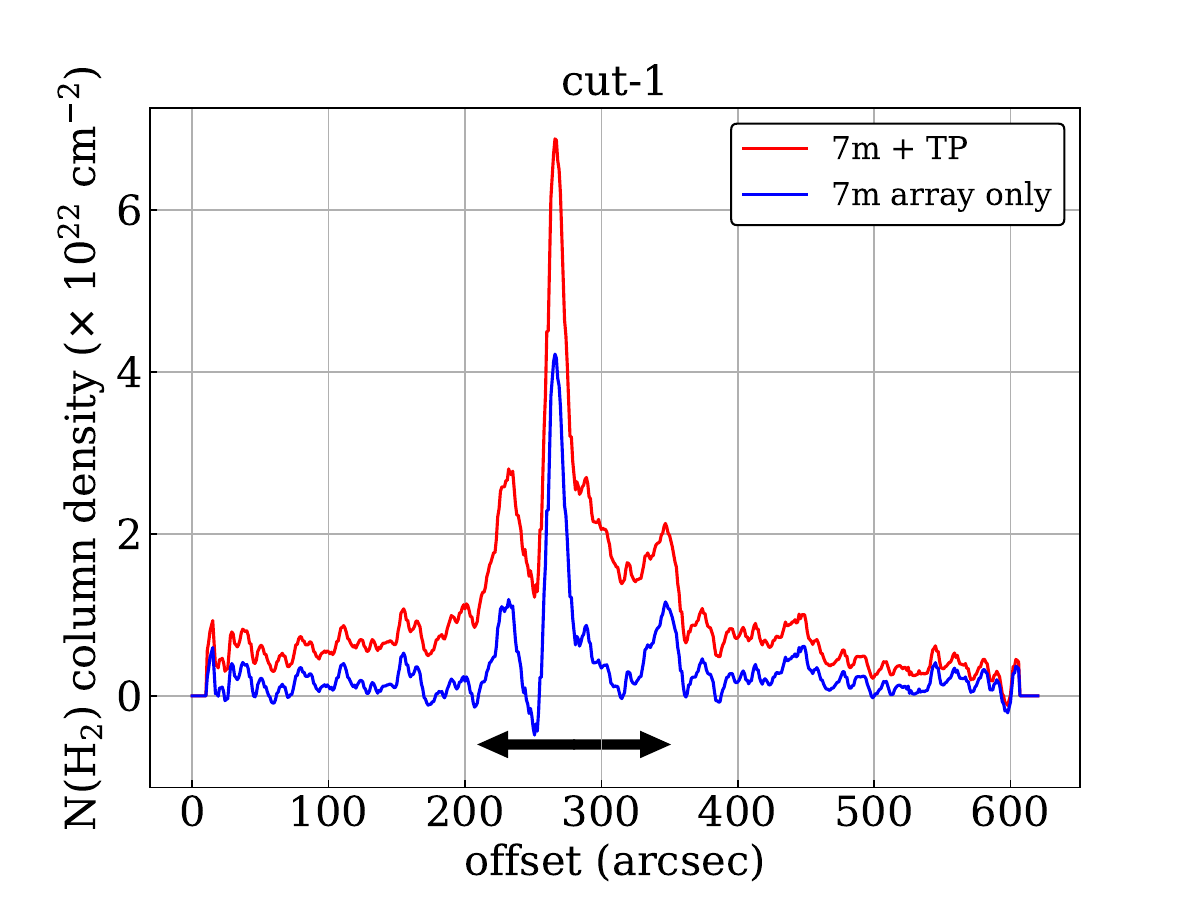}
    \includegraphics[width=8cm]{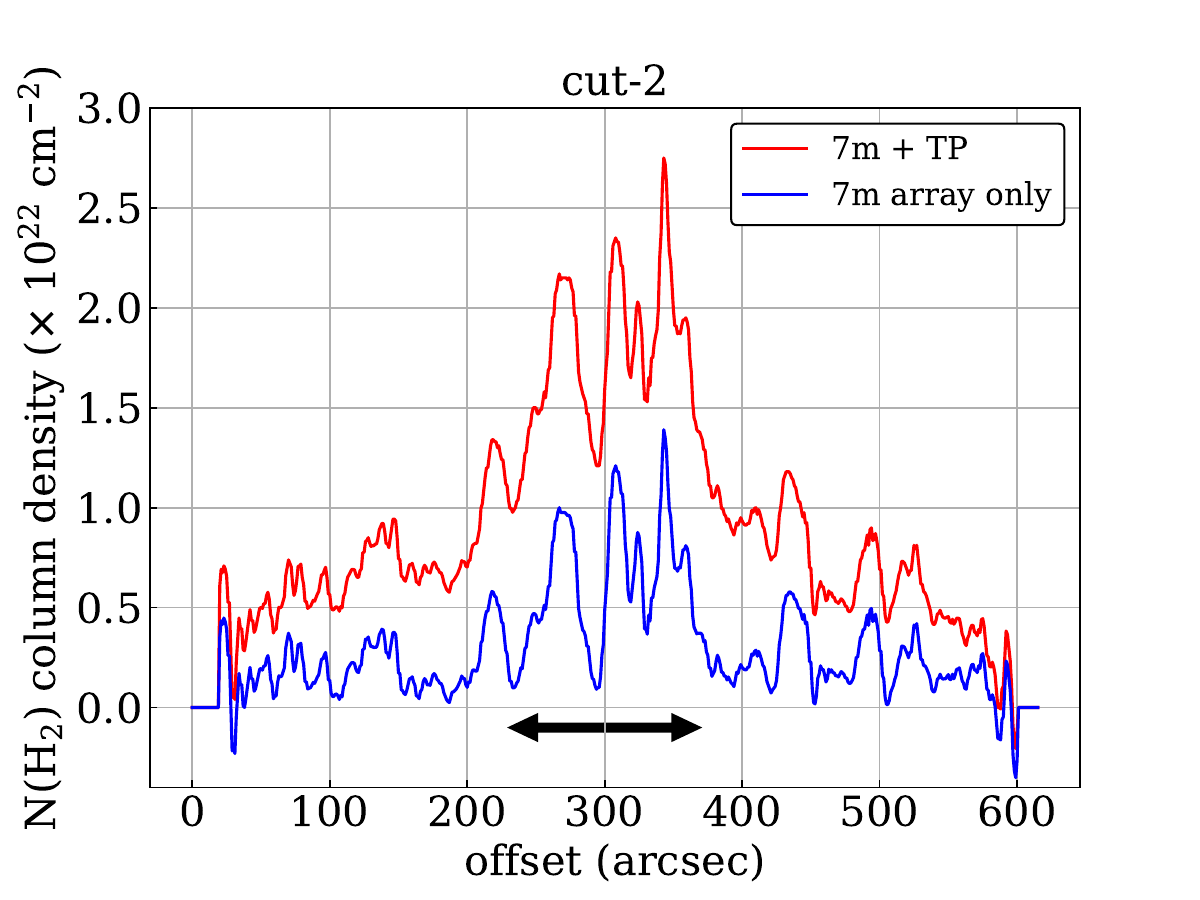}
    \includegraphics[width=8cm]{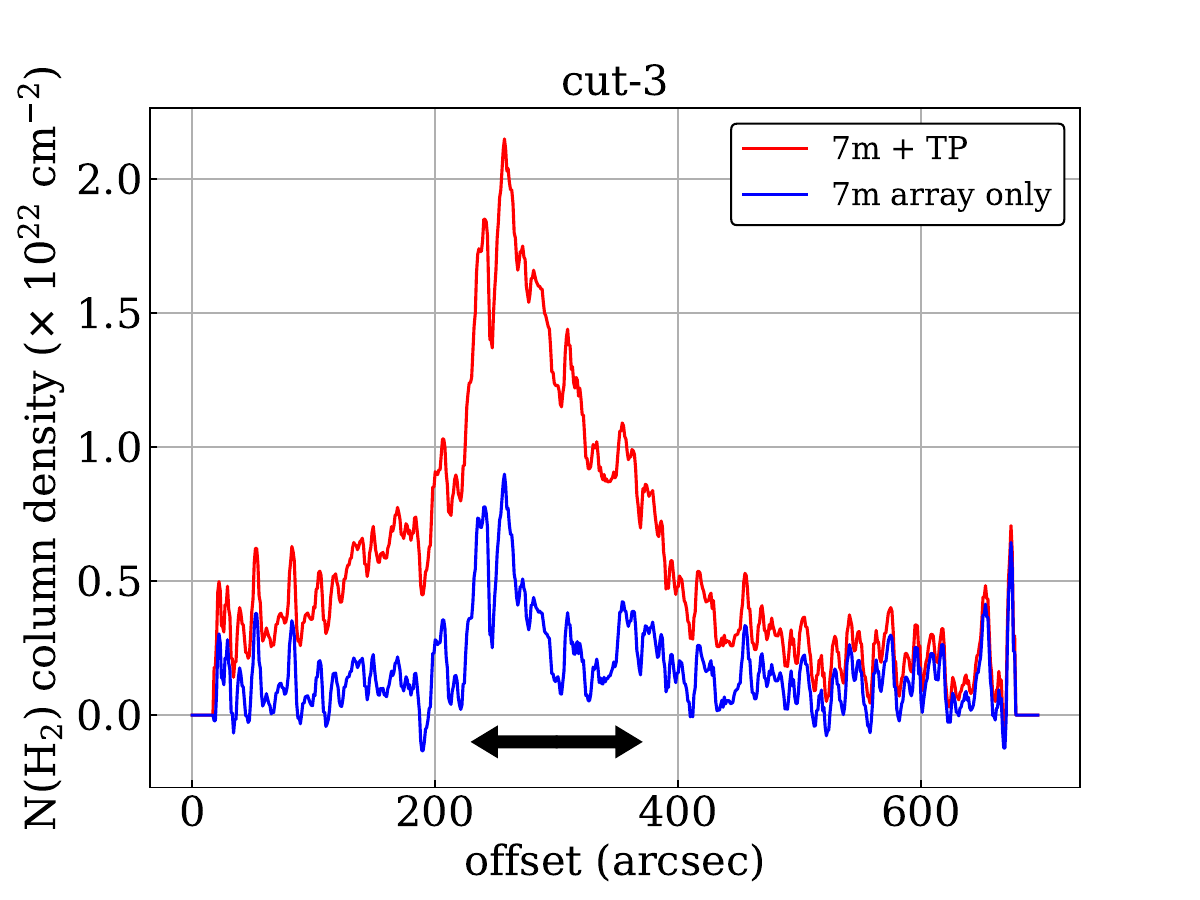}
    \includegraphics[width=8cm]{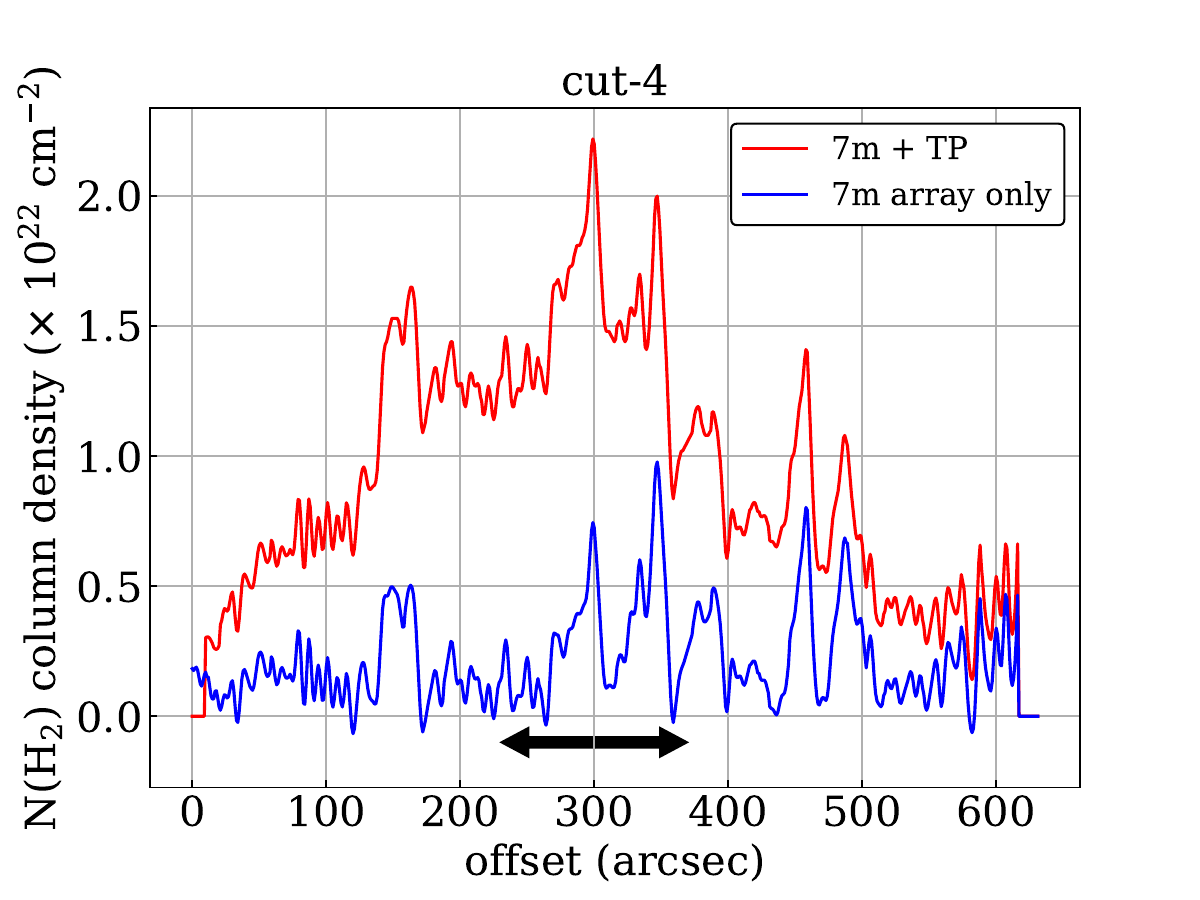}
    \caption{
    H$_2$ column densities profiles cut across the cloud at the locations indicated by the arrows in Fig.\ \ref{fig:finfinder}. The column density profiles obtained from the 7m-array only data are shown in blue, while those from the TP combined data are in red. The arrows indicate the typical filament width of 0.1 pc revealed by Herschel observations.
}
    \label{fig:column_cut}
\end{figure}

The filamentary structures are confirmed in the column density profiles of Fig.\ \ref{fig:column_cut} cut across the cloud at several locations. Those obtained with the 7m array alone are shown in blue, and those combined with the TP array are shown in red. The filamentary structures have spiky column density peaks in the combined profile, and they are distributed on top of the $\ga 0.1$ pc scale extended emission recovered by the TP array. The smoothly stretched structures are filtered out and not detected by the 7m array alone. These indicate that the sub-structures are embedded in the filaments identified by the Herschel observation, as thinner sub-filaments, and were detected for the first time by resolving the internal structure by high resolution observations. 
From now on, we refer to these thin filament-like structures as ``feathers".
\citet{hacar18} defined filamentary cloud substructures named as ``fibers", but they are qualitatively and quantitatively different from feathers as to be discussed in Section \ref{subsection:feathers}.

\begin{figure}[ht]
    \centering
    \includegraphics[width=8cm]{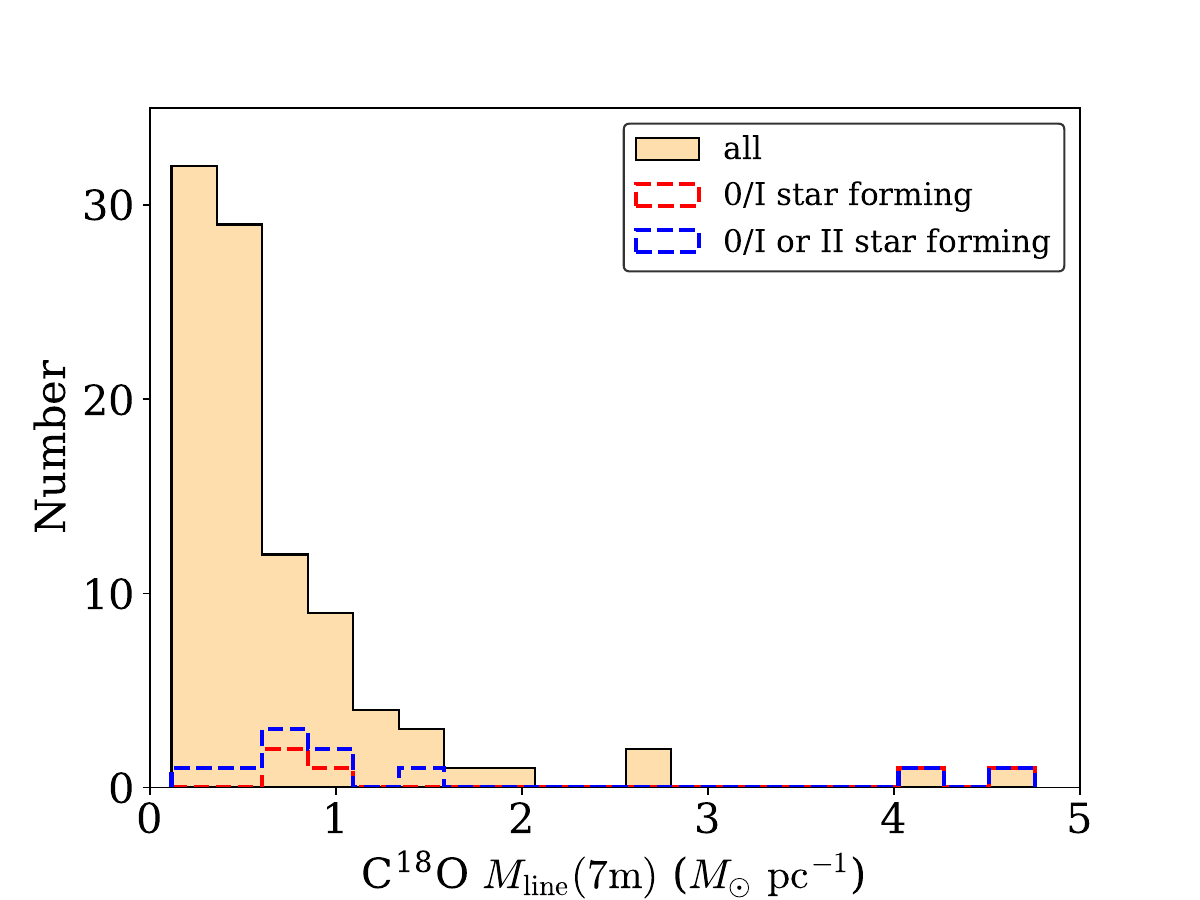}
    \includegraphics[width=8cm]{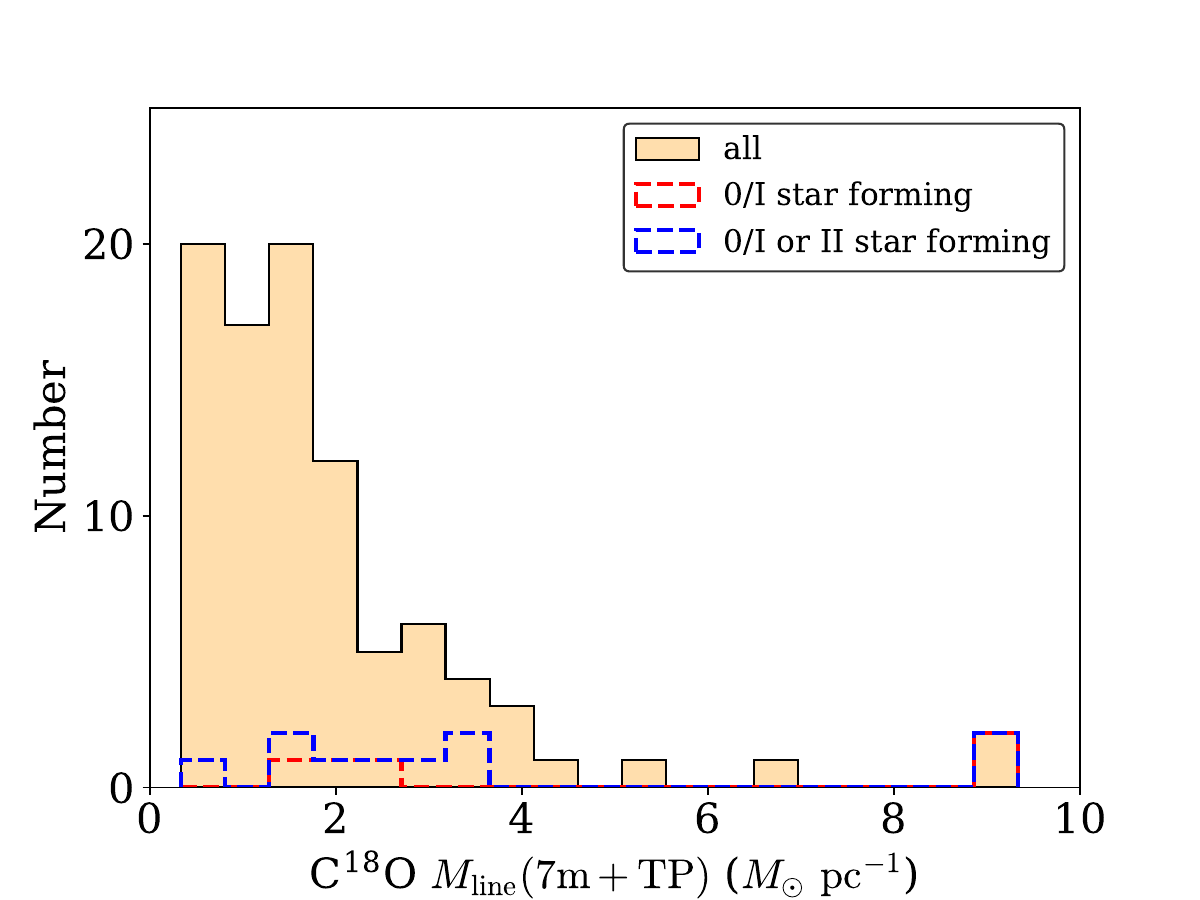}
    \caption{
    Histograms of the line mass of the C$^{18}$O feathers, $M{\rm _{line}(7m)}$ on the left and $M{\rm _{line}(7m+TP)}$ on the right. The broken lines denote the number distributions of those associated with YSOs (Class 0/I are in red, and Class 0/I+II are in blue). Note that 3 feathers are located in the region with no TP-array data, and hence no estimations of $M{\rm _{line}(7m+TP)}$.
    }
    \label{fig:linemass}
\end{figure}

For each feather identified by FilFinder, we estimated the mass per unit length (line mass, $M_{\rm line}$) from the average column density and the filament width obtained above. We use column densities at each pixel derived from two different ways, i.e., $N({\rm H_2})$ by the 7m-array only data and those by the 7m+TP-array data, and obtained $M{\rm _{line}(7m)}$ and $M{\rm _{line}(7m+TP)}$ as shown in Fig.\ \ref{fig:linemass} as histograms.
They were distributed in the range of 0.1 -- 4.8 $M_\sun\,{\rm pc}^{-1}$ with a median value of 0.5 $M_\sun\,{\rm pc}^{-1}$, and 0.3 -- 9.3 $M_\sun\,{\rm pc}^{-1}$ with a median value of 1.4 $M_\sun\,{\rm pc}^{-1}$, respectively. The critical line mass at which the filament becomes gravitationally unstable and fragments into molecular cloud cores is expressed as $2\,C_s^2/G$, where $C_s$ is the sound speed and $G$ is the gravitational constant. The critical line mass for gas with temperature of 15 K is $24\ M_{\sun}\,{\rm pc}^{-1}$, which is significantly larger than most of the feathers, while the two feathers with the largest $M{\rm _{line}(7m+TP)}$ of $\sim 9$ $M_\sun\,{\rm pc}^{-1}$ seem to be in transcritical conditions. 
On the other hand, from Fig.\ \ref{fig:column_cut}, assuming that the extended structure with a column density of $\sim 1.5\times 10^{22}$ cm$^{-2}$ at the center is a filament with the width of 0.1 pc, the line mass is approximately $30\ M_{\sun}\,{\rm pc}^{-1}$. Although the CrA cloud is likely to be under the influence of external perturbations as discussed below, and the temperature and the sound speed may increase due to the feedback effects from the young stars, the 0.1-pc width filament can be gravitationally super critical up to the temperature of 20 K.

It is difficult to determine the H$_2$ column density directly from the SO line intensity in general because the SO molecules are known to vary in their abundance depending on the environment. As the SO molecule sublimates from dust grains by gentle shocks, observed SO emission lines suggest the presence of past shock waves. Since the feather structures detected in SO show similar widths and lengths to those of C$^{18}$O, we assume the average density same as the critical density of $2 \times 10^6$ cm$^{-3}$ for the SO emission lines, and adopt the feather width of 1500 AU comparable to C$^{18}$O, the line mass of $\sim 7~M_{\sun }\,{\rm pc}^{-1}$ is obtained. This is comparable to the largest $M{\rm _{line}(7m+TP)}$ of the C$^{18}$O feathers, but still below the critical value.

\subsection{Velocity fields}

\begin{figure}[ht]
    \centering
    \includegraphics[height=7cm]{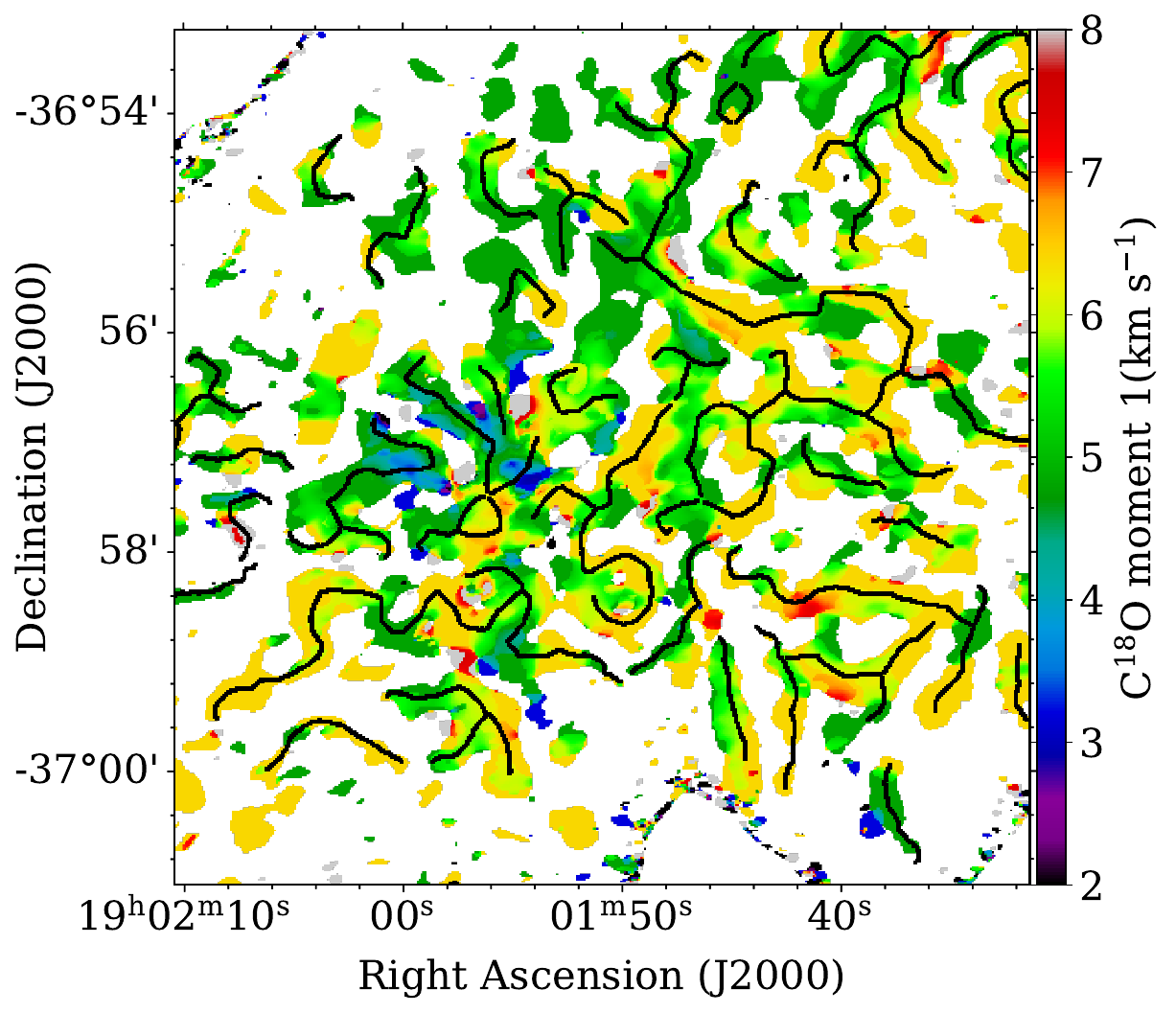}
    \caption{Distribution of the velocity field (moment 1) of the C$^{18}$O emission line detected by the 7m array, with the feather structures identified by FilFinder overlaid with black lines. Note that the velocity resolution is 1.6 km s$^{-1}$.}
    \label{fig:moment1}
\end{figure}

Figure \ref{fig:moment1} shows the moment 1 map of the C$^{18}$O line taken by the 7m array. In the present spectral setup, the velocity resolution of C$^{18}$O is 1.6 km s$^{-1}$, which is not sufficient to discuss small changes in the velocity field and spectral shapes. In spite of the poor velocity resolution, the moment 1 map exhibits noticeable velocity change among feathers. The velocity dispersion among feathers is about $\sim 1$ km s$^{-1}$. On the other hand, although the filament identification by FilFinder was performed for the two-dimensional distribution from the peak $T$ map, it can be seen that each feather is defined as a relatively coherent velocity structure. Despite the limitation of the velocity resolution, the cloud interior is expected to be highly turbulent even in the dense clump, while rather monolithic structures of feathers are embedded.

%% file: 4_Discussions.tex
\section{Discussions}
\label{section:discussions}

\subsection{Distribution of the feather structures}
\label{subsection:feathers}

As shown in Sec.\ \ref{section:filament}, the column density distribution shows that the feather structures with widths of $\sim 1000$ AU scale are embedded inside the extended structures of $\ga 0.1$ pc. 
The feathers are different in nature from the filamentary molecular clouds of about 0.1 pc in width, which have been ubiquitously discovered by observations of the Herschel and other telescopes, and whose formation mechanisms have been actively debated \citep[e.g.,][]{palmeirim13}.
As with many other molecular clouds, filamentary structures have been identified from the Herschel observations by \citet{bresnahan2018} in the R CrA cloud. Although the filaments are entangled and not well resolved in the dense head part of the cloud, the column density profile of a filament at the cloud outskirt is in good agreement with other filaments of 0.1 pc width. Feathers are therefore considered to be small-scale substructures associated with filaments of 0.1 pc width. 
In isolated star-forming regions such as Taurus, filaments fragment along the axis to form molecular cloud cores, where ongoing star formations are observed. Bundles of multiple filaments with different velocity components have also been observed in some molecular clouds \citep{mizuno95,hacar13}. However, with the similar ACA survey in Taurus by \citet{tokuda20}, no filamentary structures were detected, and only sub-structures of the prestellar cores were observed. 

Detections of filamentary structures narrower than 0.1 pc have been reported in several star forming regions. For example, in the Orion Molecular Cloud, OMC 1-3 region, ALMA observations of N$_2$H$^+$ have detected structures with a width of about 0.035 pc,  called fibers \citep{hacar18}, while feathers have typical width of 0.005 pc or 1000 AU. Many of the fibers are distributed parallel to the larger-scale elongated structure of the cloud obtained at lower resolution or radially from the dense clumps forming hub-filament structures. They have coherent velocity structure.
In contrast, the feather structures observed in the CrA cloud are randomly oriented, indicating that the internal structure of the filaments is more complex.
The typical line masses are also different by more than an order of magnitude as $M_{\rm line} \sim$ 20-35 $M_\sun\,{\rm pc}^{-1}$ for the fibers and $M_{\rm line} \sim 1~M_\sun\,{\rm pc}^{-1}$ for the feathers (see the next subsection). With respect to the filament width, the feathers are more like the CO striations detected in the $\rho$ Oph B2 region by ALMA \citep{kamazaki2019}. The CO striations have 600-1300 AU width running parallel to the magnetic fields, while their line masses are not estimated.
In addition, some feathers have complex internal velocity structures as shown by high resolution 12m array data (to be demonstrated in the part 2 paper), although the ACA data do not have high enough velocity resolution.

\begin{figure}[ht]
    \centering
    \includegraphics[height=6.2cm]{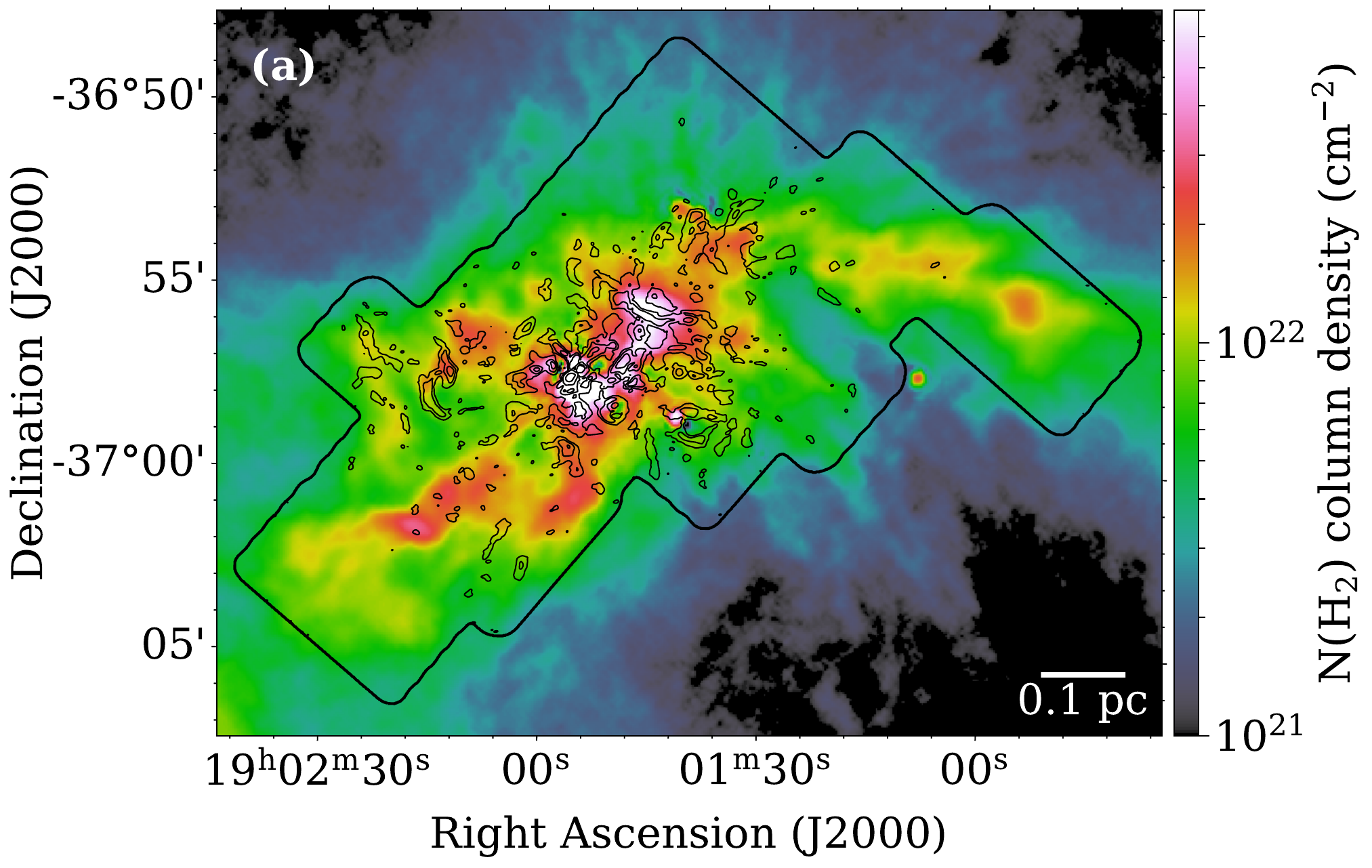}
    \includegraphics[height=6.2cm]{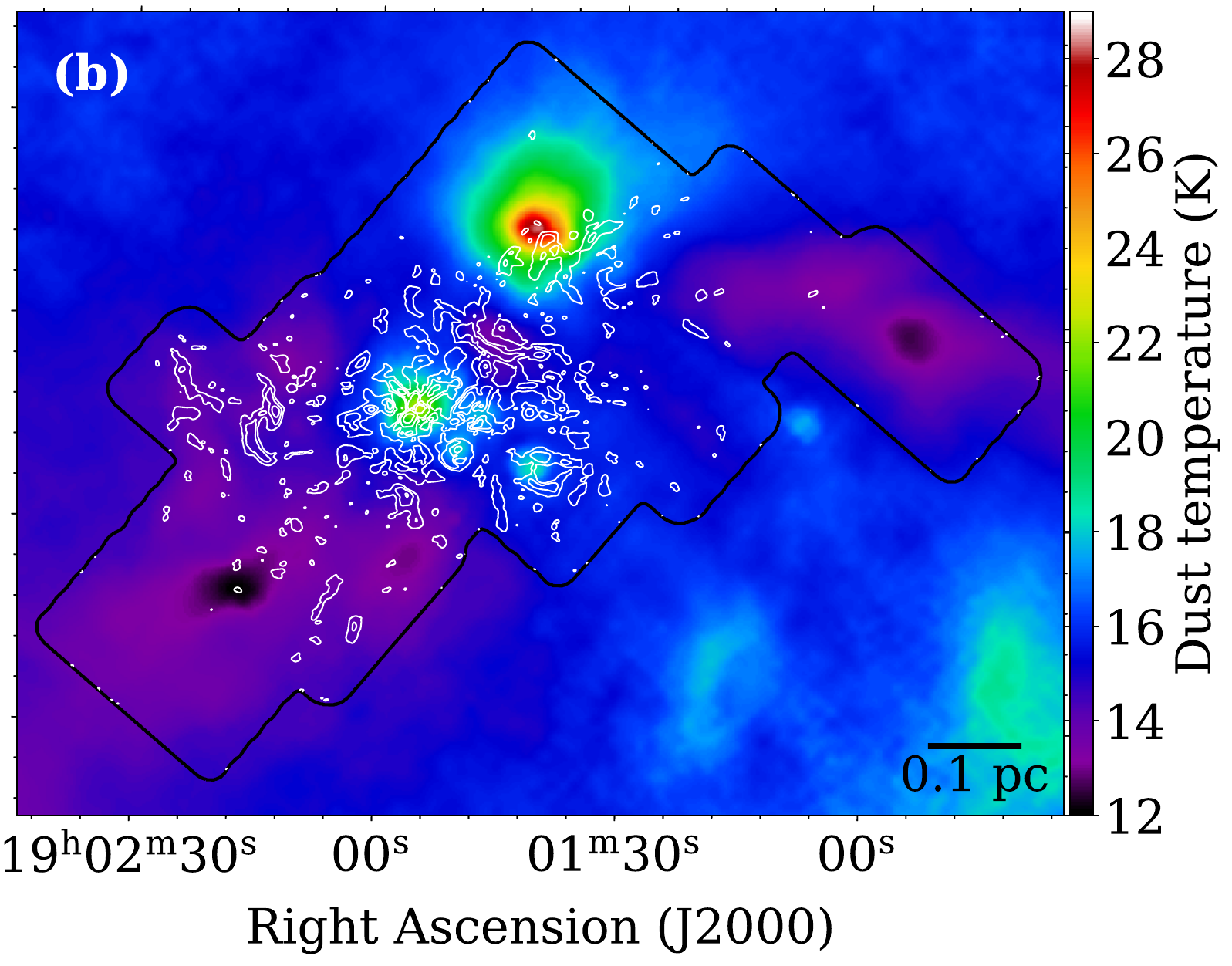}
    \includegraphics[height=6.2cm]{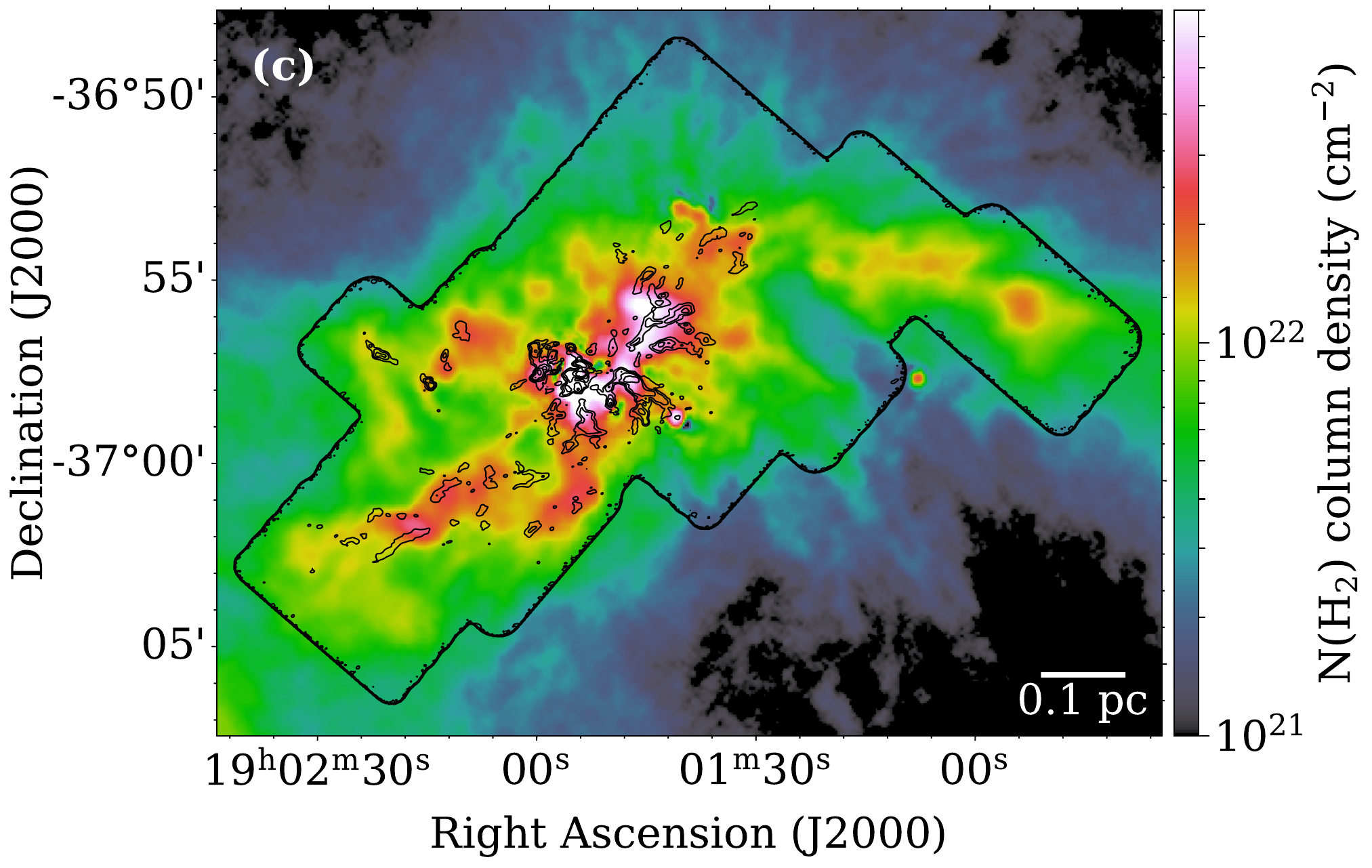}
    \includegraphics[height=6.2cm]{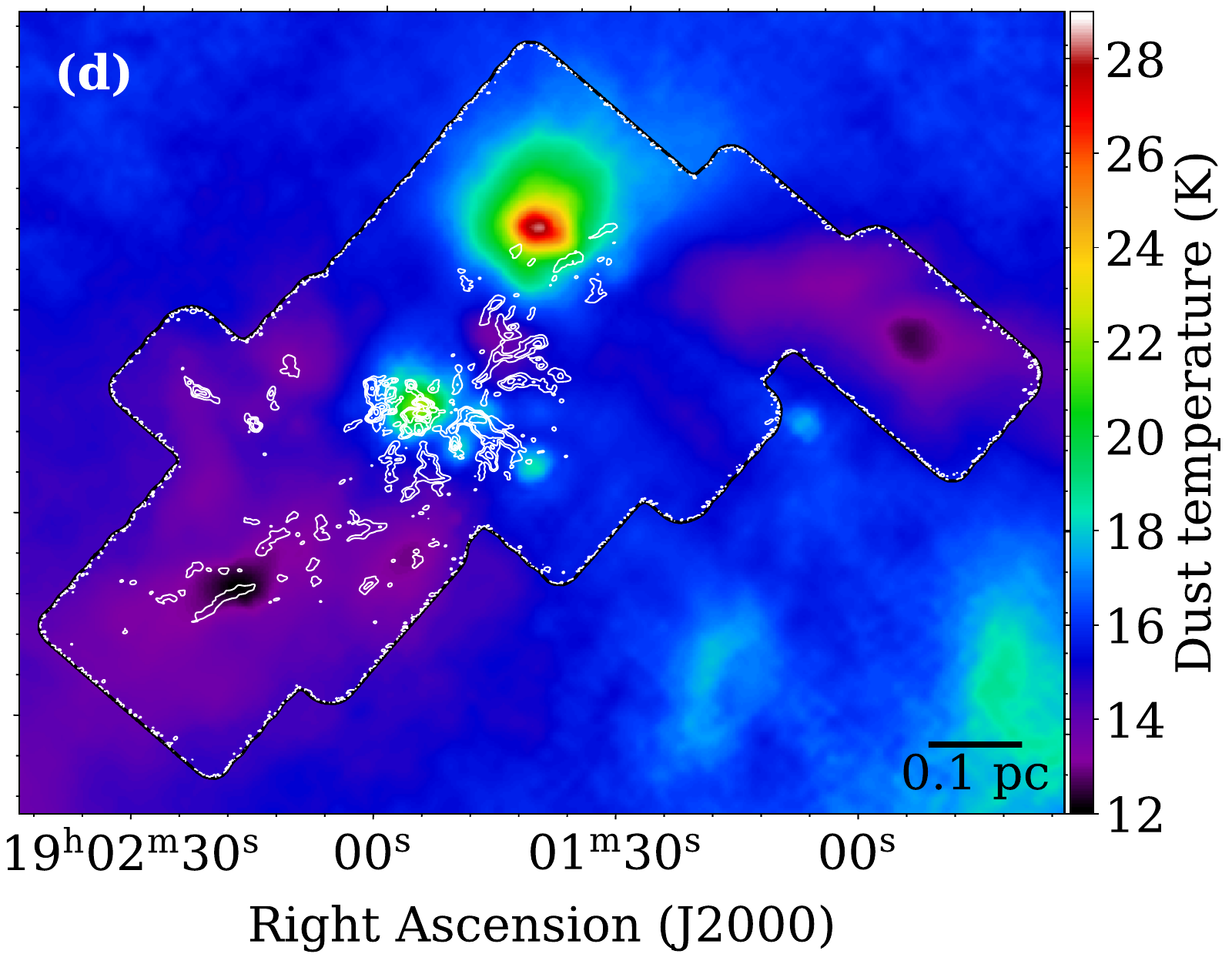}
    \caption{(a, b): The peak antenna temperature distributions of C$^{18}$O contours overlayed on the H$_2$ column density and dust temperature maps obtained from the Herschel Telescope Gould Belt Project, respectively. Contour levels are 0.25, 0.75, 1.5, and 3.5 K. (c, d): Same as (a, b), but for SO. Contour levels are 0.2, 0.5, 1.0, 2.0, and 4.0 K.}
    \label{fig:dust_col_temp}
\end{figure}

The extended large-scale gas distribution as detected by the TP-array observations tends to show increasing column density toward the cluster center, although their sensitivity are not uniform. Fig.\ \ref{fig:dust_col_temp} shows the dust column density and temperature distributions from the Herschel Gould Belt Survey \citep{andre10}, on which 7m-array peak temperature of C$^{18}$O and SO lines are superimposed with contours. 
The dust column density is particularly high near the center of the clump, where concentrated YSOs are located as the cluster, and the feather structures are also concentrated in the both emission lines. 
Besides, the C$^{18}$O feathers are distributed over lower-density regions in the periphery of the clump. 
In contrast, the SO feathers are more localized in high column density regions, as well as in the south-eastern extension of the tail along the ridge where no C$^{18}$O feather is detected. 
The dust temperature is particularly low below $\sim 13$ K in this tail, and C$^{18}$O intensity clearly decreases here. These results suggest that C$^{18}$O is depleted in high density regions away from the cluster where the dust temperature is low. However, such region is limited and a uniform abundance distribution is assumed in this paper. To derive more precise column densities, observations of N$_2$H$^+$, which is less affected by depletion and have relatively stable molecular abundances, would be required.

\subsection{Line mass and dynamical stability of the feathers}

Typically feathers have small line masses. The critical line mass assuming $T=15$ K is $24~M_{\sun}\,{\rm pc}^{-1}$, whereas medians of the observed line masses $M{\rm _{line}(7m)}$ and $M{\rm _{line}(7m+TP)}$ are $0.5~M_\sun\,{\rm pc}^{-1}$ and $1.4~M_\sun\,{\rm pc}^{-1}$, which corresponds to the kinetic temperature of below 1 K. These are extremely low values although the dispersion of the line mass is about one order of magnitude. The median density of the feather is $2.1 \times 10^5$ cm$^{-3}$, and the highest one is $1.2 \times 10^6$ cm$^{-3}$. From these facts, it seems that the feather structures cannot be unstable with self-gravity alone and do not fragment and collapse. However, external pressure or other disturbance may support them for star formation. 
On the other hand, the spatial associations of the C$^{18}$O feathers with the YSOs detected by Spitzer is weak, and only 11 out of 101 feathers have YSOs associated with them, as described in Sec.\ \ref{section:filament}. As can be seen in the histograms by broken lines in Fig.\ \ref{fig:linemass}, there is no clear trend in the observed line masses associated with star-formation. These suggest that most of the feathers are expected to be structures not forming stars but rather transient. Such high proportion of unbound structures has also been reported in Orion \citep{sato2023}.
The exceptional region is the feathers associated with concentration of 4 Class 0/I protostars including IRS 7A, 7B, CXO 34, and SMA 2. The 2 feathers with largest $M{\rm _{line}(7m+TP)}$ of $> 9~M_\sun\,{\rm pc}^{-1}$ (see Fig.\ \ref{fig:linemass} right) are the branches of a common skeleton located at the cluster center. Note that it is a four-fork skeleton and IRS 7B is located at the junction point. After careful visual inspection of high resolution 12m array data, we judged that IRS 7B is associated with the highest line mass feature, and IRS 7A, CXO 34, and SMA 2 are with the smoothly connected 2nd highest line mass feature. These are particularly interesting as the 4 young YSOs are embedded in one coherent feather separated by $\sim 1000$ AU each. This is likely the very site where cluster formation is underway by fragmentation of the feather.
Further discussion with higher spatial and velocity resolution data will be made in the part 2 paper. 

\subsection{Filament formation mechanisms}

The existence of filamentary molecular clouds has been known for a long time, and even with a resolution of slightly less than 0.1 pc, elongated structures of velocity-continuous molecular gas have been identified in the Taurus and Orion molecular clouds, with high-density cores embedded within them \citep[e.g.,][]{mizuno95, nagahama98}. Furthermore, the existence of a magnetic field running perpendicular to the filament and the existence of a striation structure of surrounding low-density gas along the field lines were known to exist \citep{goldsmith08, heyer2008, palmeirim13, matthews2014, franco2015, cox2016, kamazaki2019}. The velocity structure has also led to the acceptance of the idea that low-density gas accumulated along the striation, forming dense filaments with 0.1 pc widths. It has also been proposed that the increase in line mass causes gravitational instability, which leads to the formation of molecular cloud cores and star formation by accelerating axial fragmentations \citep{pineda2023}. Although the contraction should proceed not only in the axial direction but also in the radial direction under the gravitational instabilities, the filaments maintain the almost constant widths of 0.1 pc. This is one of the mysteries yet to be solved.

On the other hand, in Infrared Dark Clouds (IRDCs), where massive stars are forming, so-called hub-filament structures have been detected, in which a massive hub exists at the center and filaments spread radially from it. Velocity gradients along the filaments are detected, and the filaments are suggested to be channel flows of accreting gas onto the hub \citep[e.g.,][]{peretto14}. Many active researches are underway on this topic, including survey-type studies \citep[e.g.,][]{Kumar20,Tokuda23LMC}.

The R CrA cloud with the head-tail structure can be viewed as an asymmetric Hub-filament structure, but its origin is intriguing because nothing like the present feather structure has ever been detected inside the head or hub. As mentioned above, the filaments identified in the Herschel data by \citet{bresnahan2018} have a line mass of about $\sim 30~M_{\sun}\,{\rm pc}^{-1}$ and are in a trans-/super-critical state. In the head section, however, the structure of these filaments is complex and disordered. This is in contrast to the long, straight and aligned filaments in the tail section.

\subsection{Comparison with MHD simulations}

\begin{figure}[ht]
    \centering
    \includegraphics[height=7cm]{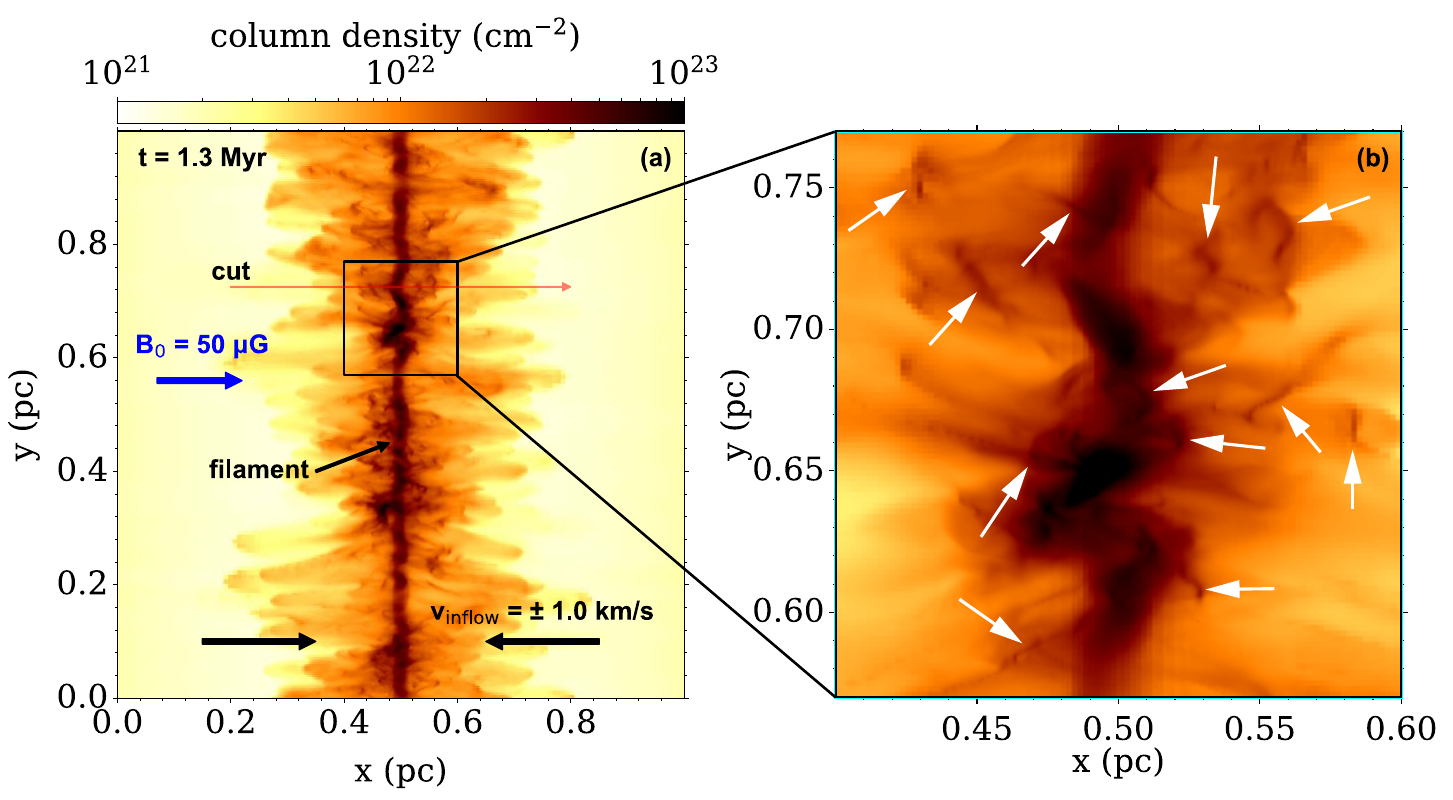}
    \includegraphics[height=6.5cm]{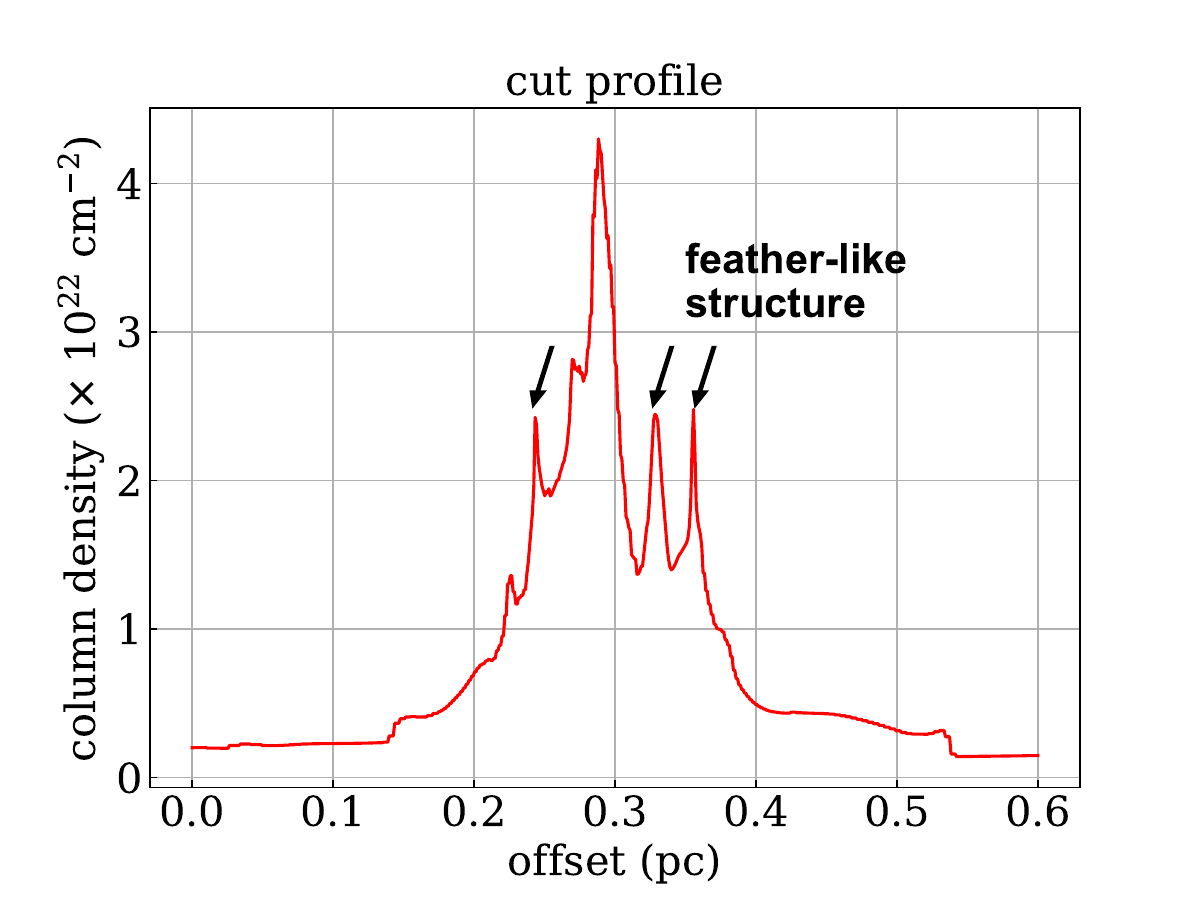}
    \caption{(top-let) Column density distribution at a snap shot of 1.3 Myr of the formed filament by the high-resolution MHD simulation (Abe et al.\ in prep.; see text for more detail). The blue arrow denotes the 50 $\mu$G initial magnetic field direction. (top-right) Zoom-up view of the central part of the simulated map where column density is the highest. The feather-like structures are visible as column density enhancements denoted by the white arrows. (bottom) a horizontally cut column density distribution.}
    \label{fig:MHDsim}
\end{figure}

To further discuss on the origin of the feather, we compare the results of magnetohydrodynamics (MHD) simulations performed at high resolution with the observed properties. Abe et al.\ (in prep.) performed MHD numerical simulations of colliding gas flows at a maximum resolution of 0.001 pc. Initially, a sheet of gas with a density of 4000 cm$^{-3}$ and thickness of 0.1 pc is placed, and a magnetic field of 50 $\mu$G penetrates in the in-plane $x$ direction. Gas flow with a velocity of $\pm 1$ km s$^{-1}$ is given from both sides in the $x$ direction, and the gas is converged at the center. Isothermal condition is assumed, adding fluctuations to the collision surface, and self-gravity and ambipolar diffusion are taken into account in the calculations. A filament is formed by the accretion flows, and the wavefront of the slow-shock is unstable as the slow-shock instability \citep{abe2024}. The instability generates wavy-fronts from the small initial fluctuations.  Gas is accumulated to the valley of the bent shock front to form dense blobs, as the gas flow across the magnetic fields by ambipolar diffusion \citep{snow2021}. 
As a result, small structures, which they call bullets, are formed and repeatedly collide with the filaments with $N({\rm H_2}) \sim 10^{23}$ cm$^{-2}$ formed on the collision surface in a reciprocating motion from both sides. The column density distribution after 1.3 Myr of calculation is shown in Fig.\ \ref{fig:MHDsim}.
The column density distribution shows a filament running vertically in the center, accompanied by fine sub-structures around it that resemble the feathers seen in the zoom-up view of Fig.\ \ref{fig:MHDsim} top-right. A horizontal cut of this column density distribution is shown in Fig.\ \ref{fig:MHDsim} bottom. Sharp density peaks with widths of about 0.01 pc are seen on top of the structure that extends about 0.1 pc. The shape of this peak is very similar to the one observed in Fig.\ \ref{fig:column_cut}, and the absolute values of the column densities are also similar. These visual characteristics reproduce the observation well.
The time evolution of the simulations shows feather-like structures in the early phase of the evolution, but they have not acquired high enough column densities to form stars. After about 1 Myr in the time evolution, large amount of gas is accumulated onto the central filament showing high column density and large line mass. 
Overall, while the filament itself is massive enough to undergo gravitational collapse, the converging flow-driven bullets act as a turbulent amplification mechanism to maintain the filament width of $\sim 0.1$\,pc. This process creates many transient $\sim 1000$ AU-scale feather-like structures as the turbulent elements. These are consistent with the observed result that many feathers are not forming stars, and dense clusters are formed near the central high line-mass filament in the CrA region. In fact, early observational studies across low- and high-mass star-forming regions confirmed velocity-linewidth enhancements as the filament column density (line mass) increases under the almost uniform 0.1\,pc width condition \citep{Arzoumanian13,Tokuda23LMC}, although the spatial resolution was not high enough to resolve the tiny substructures within the filaments. We note that one of the feather structures shows a ring-shaped morphology and is accompanied by the Class I protostar, IRS~2, at the edge. Our separate work \citep{Tokuda23Interc} pointed out that the mechanism of magnetic flux transport from the protostellar disk can also explain the formation of the ring feature \citep[see also][]{tokuda2024}. This might be an exceptional case in the region as a whole. To explain the global molecular cloud substructures, it is essential to consider the overall history of large-scale gas dynamics.

On the theoretical side, there is an increasing number of numerical simulation studies dealing with filament formation. \citet{moeckel2015} reproduced filaments whose widths are from 0.4 to 1 pc from isothermal turbulent initial condition. \citet{smith2016} demonstrated formation of bundle of subfilaments with typical radius of 0.035 pc from hydrodynamic turbulent cloud. \citet{clarke2018} used initial setup of inflowing gas in the simulation, and identified structures that mimic the fibers observed by \citet{hacar13}. However, none of these simulations take the magnetic fields into account. 
Given that the magnetic field drastically affects the structure formation in dense molecular clouds \citep{fukui21, pineda2023}, we need further theoretical studies of the filament formation under the influence of magnetic fields with ambipolar diffusion.

Below we discuss how the formation mechanism of the feathers could be explained by the colliding gas flows suggested by the MHD simulations. As a condition for the simulation, a sheet-like molecular cloud is initially prepared. According to \citet{inoue_fukui13}, the gas is first collected in a sheet-like shape by the shock wave of the first gas collision. At that time, magnetic field lines are bundled in the plane of the sheet. Such magnetic filed morphology following the \cite{inoue_fukui13} model are actually observed in another dense core at the CrA region \citep{Kandori20}. The second shock wave is generated by the continuous inflow of gas onto the filament, and bullets generated by the slow-shock instability and the ambipolar diffusion develop as mentioned above. Thus multiple or continuous gas flow for $\ga 1$ Myr is required.
The characteristic shape of the CrA cloud suggests that it is subject to external perturbations: as the head is pointing toward the Upper Sco OB association subgroup, \citet{tachihara02} suggesting that the entire cloud is affected by an expanding \ion{H}{1} shell caused by a past supernova explosion \citep{degeus92}. Furthermore, \citet{bracco20} points out that the CrA cloud is located on the boundary where two \ion{H}{1} shells overlap. They have line-of-sight velocities of 3.3 km s$^{-1}$ and 7.4 km s$^{-1}$, which is somewhat larger than the velocity difference of the colliding gases assumed in the simulation. Although no supernova remnants have been identified in these shells, such expanding gas motion is caused by stellar winds or photo-evaporation effects \citep[e.g.,][]{tachihara2000}{}. The entire cloud has extended complex morphological and velocity structures (see Fig.\ \ref{fig:NANTEN} in Appendix \ref{sec:Nanten}). 
The elongated tail of the cloud consists of multiple filamentary structures. The northern part of the tail (upper left of in each panel) appears from 3.5 km s$^{-1}$ and is strong around 5 km s$^{-1}$, while south-west part (right side in each panel) is visible from 4.5 km s$^{-1}$ to 7.5 km s$^{-1}$. These structures are at the velocity between the two expanding \ion{H}{1} shells. The head part of the cloud has a large velocity dispersion, but the large-scale structures suggest that the entire cloud is under the influence of the external disturbance.
Alternative interpretation is that the cloud is formed as an impact of high-velocity cloud falling onto the Galactic plane, as supported by stellar space motions \citep{neuhauser1998}. 
These situations are consistent with the assumption that the CrA cloud was formed by the collisional gas flow. 
The active cluster formation in the head region is potentially triggered by such a collisional gas flow. Although triggered star formation by cloud-cloud collision (CCC) has been suggested in many massive star-forming regions including the Orion Nebula Cloud (ONC) \citep[][for review]{fukui21}, intermediate-mass cluster forming regions have also been suggested to be triggered by CCCs, as NGC 1333 is considered to be a triggered SFR by CCC as well \citep{loren76}. It is suggested that gas collision with various kinetic energy (different flow velocity, density of flowing gas, and its duration) may trigger filament formation with various forms, which in turn trigger star formation on various activities. The feather structures detected in this study may be indicative of this scenario.

%% file: 5_Conclusions.tex
\section{Conclusions}
\label{section:conclusions}

Mosaic observations covering the central 175 arcmin$^2$ of the cluster-forming region the R CrA cloud have been carried out with the ALMA ACA standalone mode. In this paper, we mainly report the results of the C$^{18}$O ($J=$2--1) and SO $J_N=6_5$--$5_4$ emission lines, with the following results.
\begin{itemize}
\item Among the 11 Class 0/I YSOs cataloged by the Spitzer and SMA, 6 are detected as continuum point sources by the 7m array observations. 
\item Observations of the 7m array reveal very thin filamentary structures with complex distribution in both the C$^{18}$O and SO lines. The observations of the TP array detect extended emissions over a scale of about 0.1 pc, corresponding to the dust filaments detected by the Herschel telescope. The smaller filamentary sub-structures are embedded inside the filaments.
\item We identified 101 and 37 of these filamentary structures in C$^{18}$O and SO, respectively, called as {\it feathers}, from the 7m array data using the FilFinder algorithm. Although only marginally resolved with respect to the beam size, their widths are approximately a few thousand AU, suggesting that they were formed by a mechanism different from that previously suggested for filamentary molecular clouds.
\item The line masses of the feathers are a few $M_\sun$ pc$^{-1}$, significantly smaller than the critical value, and only about 10 feathers are associated with YSOs, suggesting that these structures are likely to be transient rather than gravitationally unstable and collapse to form stars. 
\item Compared to the results of MHD simulations of collisional gas flows performed at high resolution, the structure and column density obtained in this observation are well reproduced. In this calculation, such structures are created on a timescale of about 1 Myr by slow-shock waves and ambipolar diffusion associated with the gas flow. They seem to correspond to the feathers and appear to be transient structures.
\item As the R CrA cloud is located at the edge where the two \ion{H}{1} shells overlap, the cloud is likely to be under the influence of collisional gas flow. Compared to other massive cluster forming regions triggered by energetic cloud-cloud collisions, it is suggested that moderate colliding gas flow may be responsible for intermediate-mass cluster formation. This idea is supported by the facts that the interiors of molecular clump is turbulent and the large-scale cloud morphology has a head-tail structure.
\end{itemize}

\begin{acknowledgments}
This research has made use of data from the Herschel Gould Belt survey (HGBS) project (\verb|http://gouldbelt-herschel.cea.fr|). The HGBS is a Herschel Key Programme jointly carried out by SPIRE Specialist Astronomy Group 3 (SAG 3), scientists of several institutes in the PACS Consortium (CEA Saclay, INAF-IFSI Rome and INAF-Arcetri, KU Leuven, MPIA Heidelberg), and scientists of the Herschel Science Center (HSC). We thank Dr.\ Yoshinori Yonekura for providing the unpublished CO ($J=$1--0) data by the NANTEN telescope. 
This paper makes use of the following ALMA data: ADS/JAO.ALMA\#2018.A.00056.S. ALMA is a partnership between ESO (representing its member states), NSF (USA) and NINS (Japan), together with NRC (Canada), MOST and ASIAA (Taiwan), and KASI (Republic of Korea), in cooperation with the Republic of Chile. The Joint ALMA Observatory is operated by ESO, AUI/NRAO and NAOJ. This work was supported by JSPS KAKENHI grant Nos.\ JP20H01945, JP21H00040, JP22H00152, JP20H05645, JP21H00049, and JP21K13962, and also by Overseas research support grant of Yamada Science Foundation.
\end{acknowledgments}

%% file: 6_Appendix.tex
\appendix
\section{Large-scale CO data of the entire CrA cloud by NANTEN}
\label{sec:Nanten}

\begin{figure}[ht]
    \centering
    \includegraphics[width=\textwidth]{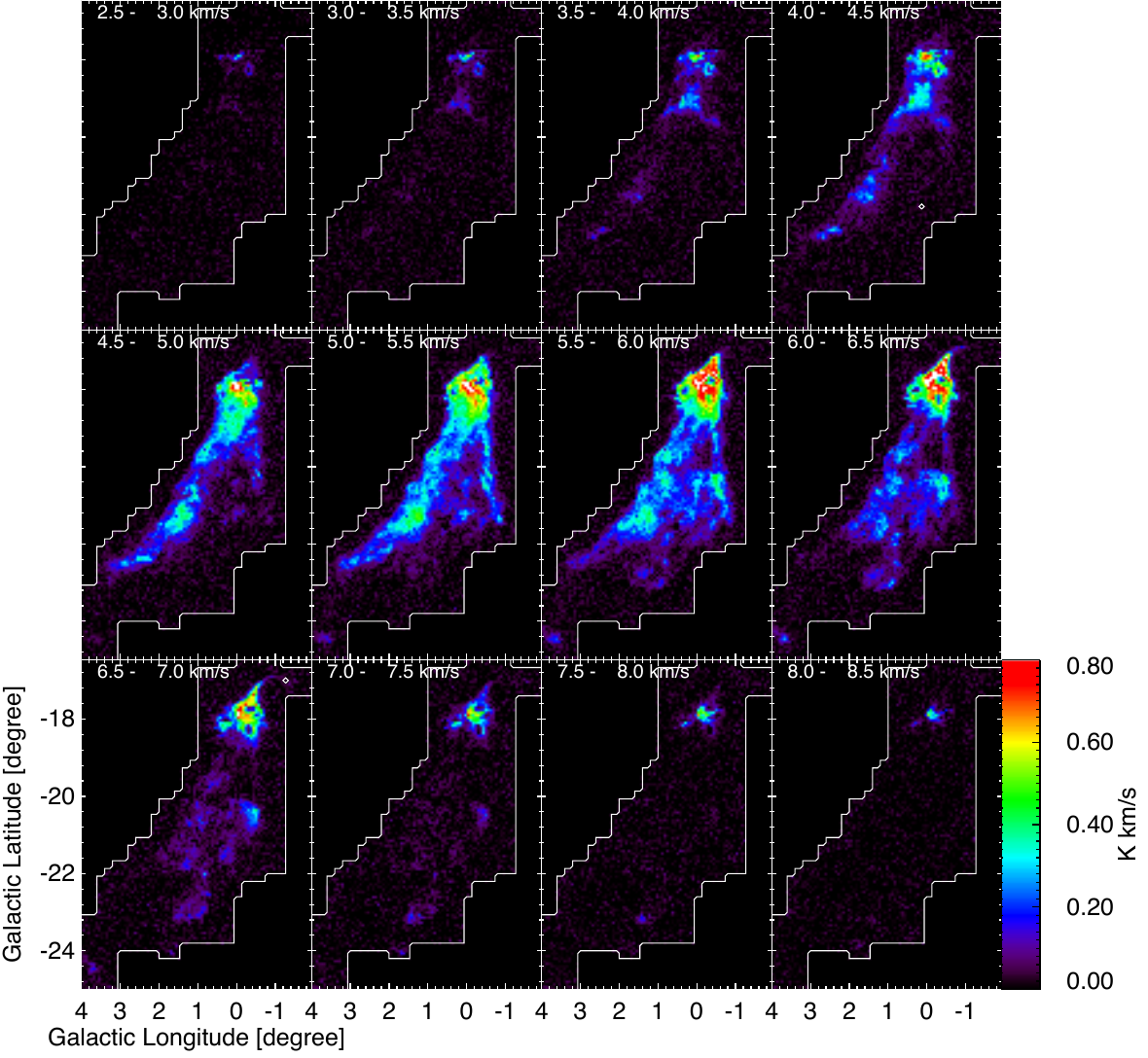}
    \caption{Channel maps of the $^{12}$CO $J=$1--0 line observed by NANTEN but yet unpublished. Each panel shows integrated intensity distribution over 0.5 km s$^{-1}$ velocity range indicated in the upper left. The while lines delineate the boundary of the observed area. 
    }
    \label{fig:NANTEN}
\end{figure}

The entire CrA cloud has been observed in $^{12}$CO $J=$1--0 line with the NANTEN telescope with the 4m dish and 2.6 arcmin half-power beam width (HPBW) resolution at 115.27120 GHz. The data were obtained in the position switching mode with 4-arcmin grid spacings and the velocity resolution of 0.1 km s$^{-1}$, but yet unpublished.
As demonstrated by the channel map of Fig.\ \ref{fig:NANTEN}, it exhibits particular velocity structures. The head part of the cloud has large velocity dispersion of 5.2 km s$^{-1}$ centered at 6.0 km s$^{-1}$, while the long-stretched tail has $\sim 1$ km s$^{-1}$ with gradient and multiple components. The eastern side of the tail has a sharp edge of intensity distribution and relatively blue-shifted, while the western side has more complex morphology and red-shifted. These imply external disturbance, possibly originated from the two \ion{H}{1} shells reported by \citet{bracco20}.

\section{Enlarged view of the central part of the cloud}
\label{sec:zoom-up}

\begin{figure}[ht]
    \centering
    \includegraphics[width=11cm]{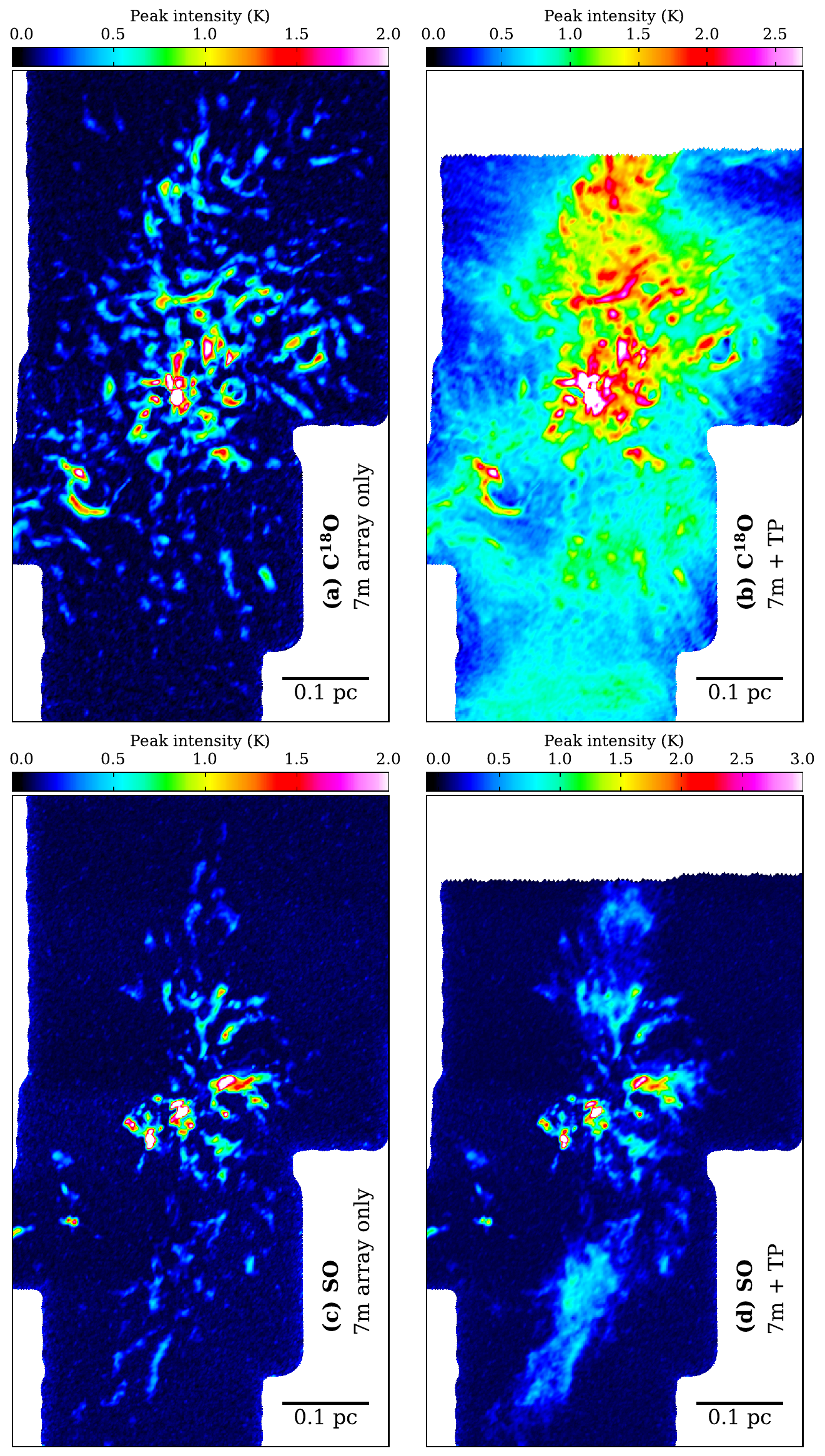}
    \caption{Zoom-up view of the central $10 \times 17$ arcmin$^2$ fields of Fig.\ \ref{fig:peakT_map} for better visualization. The images are tilted by 40 deg in counterclockwise. 
    }
    \label{fig:mom0rotate}
\end{figure}

Fig.\ \ref{fig:mom0rotate} shows enlarged peak $T$ maps of the $10 \times 17$ arcmin$^2$ central parts of the observed field where crowded small-scale structures are detected. 

%% file: main.bbl
\begin{thebibliography}{}
\expandafter\ifx\csname natexlab\endcsname\relax\def\natexlab#1{#1}\fi
\providecommand{\url}[1]{\href{#1}{#1}}
\providecommand{\dodoi}[1]{doi:~\href{http://doi.org/#1}{\nolinkurl{#1}}}
\providecommand{\doeprint}[1]{\href{http://ascl.net/#1}{\nolinkurl{http://ascl.net/#1}}}
\providecommand{\doarXiv}[1]{\href{https://arxiv.org/abs/#1}{\nolinkurl{https://arxiv.org/abs/#1}}}

\bibitem[{{Abe} {et~al.}(2024){Abe}, {Inoue}, \& {Shu-ichiro}}]{abe2024}
{Abe}, D., {Inoue}, T., \& {Shu-ichiro}, I. 2024, \apj, 961, 100,
  \dodoi{10.3847/1538-4357/ad072a}

\bibitem[{{Andr{\'e}} {et~al.}(2014){Andr{\'e}}, {Di Francesco},
  {Ward-Thompson}, {Inutsuka}, {Pudritz}, \& {Pineda}}]{andre14}
{Andr{\'e}}, P., {Di Francesco}, J., {Ward-Thompson}, D., {et~al.} 2014, in
  Protostars and Planets VI, ed. H.~{Beuther}, R.~S. {Klessen}, C.~P.
  {Dullemond}, \& T.~{Henning}, 27--51,
  \dodoi{10.2458/azu_uapress_9780816531240-ch002}

\bibitem[{{Andr{\'e}} {et~al.}(2010){Andr{\'e}}, {Men'shchikov}, {Bontemps},
  {K{\"o}nyves}, {Motte}, {Schneider}, {Didelon}, {Minier}, {Saraceno},
  {Ward-Thompson}, {di Francesco}, {White}, {Molinari}, {Testi}, {Abergel},
  {Griffin}, {Henning}, {Royer}, {Mer{\'\i}n}, {Vavrek}, {Attard},
  {Arzoumanian}, {Wilson}, {Ade}, {Aussel}, {Baluteau}, {Benedettini},
  {Bernard}, {Blommaert}, {Cambr{\'e}sy}, {Cox}, {di Giorgio}, {Hargrave},
  {Hennemann}, {Huang}, {Kirk}, {Krause}, {Launhardt}, {Leeks}, {Le Pennec},
  {Li}, {Martin}, {Maury}, {Olofsson}, {Omont}, {Peretto}, {Pezzuto}, {Prusti},
  {Roussel}, {Russeil}, {Sauvage}, {Sibthorpe}, {Sicilia-Aguilar}, {Spinoglio},
  {Waelkens}, {Woodcraft}, \& {Zavagno}}]{andre10}
{Andr{\'e}}, P., {Men'shchikov}, A., {Bontemps}, S., {et~al.} 2010, \aap, 518,
  L102, \dodoi{10.1051/0004-6361/201014666}

\bibitem[{{Arzoumanian} {et~al.}(2013){Arzoumanian}, {Andr{\'e}}, {Peretto}, \&
  {K{\"o}nyves}}]{Arzoumanian13}
{Arzoumanian}, D., {Andr{\'e}}, P., {Peretto}, N., \& {K{\"o}nyves}, V. 2013,
  \aap, 553, A119, \dodoi{10.1051/0004-6361/201220822}

\bibitem[{{Arzoumanian} {et~al.}(2011){Arzoumanian}, {Andr{\'e}}, {Didelon},
  {K{\"o}nyves}, {Schneider}, {Men'shchikov}, {Sousbie}, {Zavagno}, {Bontemps},
  {di Francesco}, {Griffin}, {Hennemann}, {Hill}, {Kirk}, {Martin}, {Minier},
  {Molinari}, {Motte}, {Peretto}, {Pezzuto}, {Spinoglio}, {Ward-Thompson},
  {White}, \& {Wilson}}]{arzoumanian11}
{Arzoumanian}, D., {Andr{\'e}}, P., {Didelon}, P., {et~al.} 2011, \aap, 529,
  L6, \dodoi{10.1051/0004-6361/201116596}

\bibitem[{{Bracco} {et~al.}(2020){Bracco}, {Bresnahan}, {Palmeirim},
  {Arzoumanian}, {Andr{\'e}}, {Ward-Thompson}, \& {Marchal}}]{bracco20}
{Bracco}, A., {Bresnahan}, D., {Palmeirim}, P., {et~al.} 2020, \aap, 644, A5,
  \dodoi{10.1051/0004-6361/202039282}

\bibitem[{{Bresnahan} {et~al.}(2018){Bresnahan}, {Ward-Thompson}, {Kirk},
  {Pattle}, {Eyres}, {White}, {K{\"o}nyves}, {Men'shchikov}, {Andr{\'e}},
  {Schneider}, {Di Francesco}, {Arzoumanian}, {Benedettini}, {Ladjelate},
  {Palmeirim}, {Bracco}, {Molinari}, {Pezzuto}, \& {Spinoglio}}]{bresnahan2018}
{Bresnahan}, D., {Ward-Thompson}, D., {Kirk}, J.~M., {et~al.} 2018, \aap, 615,
  A125, \dodoi{10.1051/0004-6361/201730515}

\bibitem[{{Bressert} {et~al.}(2010){Bressert}, {Bastian}, {Gutermuth},
  {Megeath}, {Allen}, {Evans}, {Rebull}, {Hatchell}, {Johnstone}, {Bourke},
  {Cieza}, {Harvey}, {Merin}, {Ray}, \& {Tothill}}]{bressert2010}
{Bressert}, E., {Bastian}, N., {Gutermuth}, R., {et~al.} 2010, \mnras, 409,
  L54, \dodoi{10.1111/j.1745-3933.2010.00946.x}

\bibitem[{{Cambr{\'e}sy}(1999)}]{cambresy1999}
{Cambr{\'e}sy}, L. 1999, \aap, 345, 965,
  \dodoi{10.48550/arXiv.astro-ph/9903149}

\bibitem[{{Cappa de Nicolau} \& {Poppel}(1991)}]{cappa91}
{Cappa de Nicolau}, C.~E., \& {Poppel}, W.~G.~L. 1991, \aaps, 88, 615

\bibitem[{{Carpenter}(2000)}]{carpenter2000}
{Carpenter}, J.~M. 2000, \aj, 120, 3139, \dodoi{10.1086/316845}

\bibitem[{{CASA Team} {et~al.}(2022){CASA Team}, {Bean}, {Bhatnagar}, {Castro},
  {Donovan Meyer}, {Emonts}, {Garcia}, {Garwood}, {Golap}, {Gonzalez Villalba},
  {Harris}, {Hayashi}, {Hoskins}, {Hsieh}, {Jagannathan}, {Kawasaki},
  {Keimpema}, {Kettenis}, {Lopez}, {Marvil}, {Masters}, {McNichols},
  {Mehringer}, {Miel}, {Moellenbrock}, {Montesino}, {Nakazato}, {Ott}, {Petry},
  {Pokorny}, {Raba}, {Rau}, {Schiebel}, {Schweighart}, {Sekhar}, {Shimada},
  {Small}, {Steeb}, {Sugimoto}, {Suoranta}, {Tsutsumi}, {van Bemmel},
  {Verkouter}, {Wells}, {Xiong}, {Szomoru}, {Griffith}, {Glendenning}, \&
  {Kern}}]{casa22}
{CASA Team}, {Bean}, B., {Bhatnagar}, S., {et~al.} 2022, \pasp, 134, 114501,
  \dodoi{10.1088/1538-3873/ac9642}

\bibitem[{{Clarke} {et~al.}(2018){Clarke}, {Whitworth}, {Spowage},
  {Duarte-Cabral}, {Suri}, {Jaffa}, {Walch}, \& {Clark}}]{clarke2018}
{Clarke}, S.~D., {Whitworth}, A.~P., {Spowage}, R.~L., {et~al.} 2018, \mnras,
  479, 1722, \dodoi{10.1093/mnras/sty1675}

\bibitem[{{Cox} {et~al.}(2016){Cox}, {Arzoumanian}, {Andr{\'e}}, {Rygl},
  {Prusti}, {Men'shchikov}, {Royer}, {K{\'o}sp{\'a}l}, {Palmeirim}, {Ribas},
  {K{\"o}nyves}, {Bernard}, {Schneider}, {Bontemps}, {Merin}, {Vavrek}, {Alves
  de Oliveira}, {Didelon}, {Pilbratt}, \& {Waelkens}}]{cox2016}
{Cox}, N.~L.~J., {Arzoumanian}, D., {Andr{\'e}}, P., {et~al.} 2016, \aap, 590,
  A110, \dodoi{10.1051/0004-6361/201527068}

\bibitem[{{de Geus}(1992)}]{degeus92}
{de Geus}, E.~J. 1992, \aap, 262, 258

\bibitem[{{Dunham} {et~al.}(2016){Dunham}, {Offner}, {Pineda}, {Bourke},
  {Tobin}, {Arce}, {Chen}, {Di Francesco}, {Johnstone}, {Lee}, {Myers},
  {Price}, {Sadavoy}, \& {Schnee}}]{dunham16}
{Dunham}, M.~M., {Offner}, S. S.~R., {Pineda}, J.~E., {et~al.} 2016, \apj, 823,
  160, \dodoi{10.3847/0004-637X/823/2/160}

\bibitem[{{Franco} \& {Alves}(2015)}]{franco2015}
{Franco}, G.~A.~P., \& {Alves}, F.~O. 2015, \apj, 807, 5,
  \dodoi{10.1088/0004-637X/807/1/5}

\bibitem[{{Frerking} {et~al.}(1982){Frerking}, {Langer}, \&
  {Wilson}}]{frerking82}
{Frerking}, M.~A., {Langer}, W.~D., \& {Wilson}, R.~W. 1982, \apj, 262, 590,
  \dodoi{10.1086/160451}

\bibitem[{{Fujishiro} {et~al.}(2020){Fujishiro}, {Tokuda}, {Tachihara},
  {Takashima}, {Fukui}, {Zahorecz}, {Saigo}, {Matsumoto}, {Tomida}, {Machida},
  {Inutsuka}, {Andr{\'e}}, {Kawamura}, \& {Onishi}}]{fujishiro20}
{Fujishiro}, K., {Tokuda}, K., {Tachihara}, K., {et~al.} 2020, \apjl, 899, L10,
  \dodoi{10.3847/2041-8213/ab9ca8}

\bibitem[{{Fukui} {et~al.}(2021){Fukui}, {Habe}, {Inoue}, {Enokiya}, \&
  {Tachihara}}]{fukui21}
{Fukui}, Y., {Habe}, A., {Inoue}, T., {Enokiya}, R., \& {Tachihara}, K. 2021,
  \pasj, 73, S1, \dodoi{10.1093/pasj/psaa103}

\bibitem[{{Fukui} {et~al.}(2019){Fukui}, {Tokuda}, {Saigo}, {Harada},
  {Tachihara}, {Tsuge}, {Inoue}, {Torii}, {Nishimura}, {Zahorecz}, {Nayak},
  {Meixner}, {Minamidani}, {Kawamura}, {Mizuno}, {Indebetouw}, {Sewi{\l}o},
  {Madden}, {Galametz}, {Lebouteiller}, {Chen}, \& {Onishi}}]{Fukui19}
{Fukui}, Y., {Tokuda}, K., {Saigo}, K., {et~al.} 2019, \apj, 886, 14,
  \dodoi{10.3847/1538-4357/ab4900}

\bibitem[{{Galli} {et~al.}(2020){Galli}, {Bouy}, {Olivares}, {Miret-Roig},
  {Sarro}, {Barrado}, {Berihuete}, \& {Brandner}}]{galli2020}
{Galli}, P.~A.~B., {Bouy}, H., {Olivares}, J., {et~al.} 2020, \aap, 634, A98,
  \dodoi{10.1051/0004-6361/201936708}

\bibitem[{{Goldsmith} {et~al.}(2008){Goldsmith}, {Heyer}, {Narayanan}, {Snell},
  {Li}, \& {Brunt}}]{goldsmith08}
{Goldsmith}, P.~F., {Heyer}, M., {Narayanan}, G., {et~al.} 2008, \apj, 680,
  428, \dodoi{10.1086/587166}

\bibitem[{{Groppi} {et~al.}(2007){Groppi}, {Hunter}, {Blundell}, \&
  {Sandell}}]{groppi07}
{Groppi}, C.~E., {Hunter}, T.~R., {Blundell}, R., \& {Sandell}, G. 2007, \apj,
  670, 489, \dodoi{10.1086/521875}

\bibitem[{{Groppi} {et~al.}(2004){Groppi}, {Kulesa}, {Walker}, \&
  {Martin}}]{groppi2004}
{Groppi}, C.~E., {Kulesa}, C., {Walker}, C., \& {Martin}, C.~L. 2004, \apj,
  612, 946, \dodoi{10.1086/422168}

\bibitem[{{Hacar} {et~al.}(2018){Hacar}, {Tafalla}, {Forbrich}, {Alves},
  {Meingast}, {Grossschedl}, \& {Teixeira}}]{hacar18}
{Hacar}, A., {Tafalla}, M., {Forbrich}, J., {et~al.} 2018, \aap, 610, A77,
  \dodoi{10.1051/0004-6361/201731894}

\bibitem[{{Hacar} {et~al.}(2013){Hacar}, {Tafalla}, {Kauffmann}, \&
  {Kov{\'a}cs}}]{hacar13}
{Hacar}, A., {Tafalla}, M., {Kauffmann}, J., \& {Kov{\'a}cs}, A. 2013, \aap,
  554, A55, \dodoi{10.1051/0004-6361/201220090}

\bibitem[{{Hamaguchi} {et~al.}(2005){Hamaguchi}, {Corcoran}, {Petre}, {White},
  {Stelzer}, {Nedachi}, {Kobayashi}, \& {Tokunaga}}]{hamaguchi2005}
{Hamaguchi}, K., {Corcoran}, M.~F., {Petre}, R., {et~al.} 2005, \apj, 623, 291,
  \dodoi{10.1086/428434}

\bibitem[{{Harju} {et~al.}(1993){Harju}, {Haikala}, {Mattila}, {Mauersberger},
  {Booth}, \& {Nordh}}]{harju93}
{Harju}, J., {Haikala}, L.~K., {Mattila}, K., {et~al.} 1993, \aap, 278, 569

\bibitem[{{Herbst} \& {Shevchenko}(1999)}]{herbst99}
{Herbst}, W., \& {Shevchenko}, V.~S. 1999, \aj, 118, 1043,
  \dodoi{10.1086/300966}

\bibitem[{{Heyer} {et~al.}(2008){Heyer}, {Gong}, {Ostriker}, \&
  {Brunt}}]{heyer2008}
{Heyer}, M., {Gong}, H., {Ostriker}, E., \& {Brunt}, C. 2008, \apj, 680, 420,
  \dodoi{10.1086/587510}

\bibitem[{{Inoue} \& {Fukui}(2013)}]{inoue_fukui13}
{Inoue}, T., \& {Fukui}, Y. 2013, \apjl, 774, L31,
  \dodoi{10.1088/2041-8205/774/2/L31}

\bibitem[{{Inutsuka} \& {Miyama}(1997)}]{inutsuka97}
{Inutsuka}, S.-i., \& {Miyama}, S.~M. 1997, \apj, 480, 681,
  \dodoi{10.1086/303982}

\bibitem[{{Kamazaki} {et~al.}(2019){Kamazaki}, {Nakamura}, {Kawabe}, {Hara},
  {Takakuwa}, {Hirano}, {Di Francesco}, {Friesen}, \& {Tamura}}]{kamazaki2019}
{Kamazaki}, T., {Nakamura}, F., {Kawabe}, R., {et~al.} 2019, \apj, 871, 86,
  \dodoi{10.3847/1538-4357/aaf857}

\bibitem[{{Kandori} {et~al.}(2020){Kandori}, {Tamura}, {Saito}, {Tomisaka},
  {Matsumoto}, {Tazaki}, {Nagata}, {Kusakabe}, {Nakajima}, {Kwon}, {Nagayama},
  \& {Tatematsu}}]{Kandori20}
{Kandori}, R., {Tamura}, M., {Saito}, M., {et~al.} 2020, \apj, 900, 20,
  \dodoi{10.3847/1538-4357/abaab3}

\bibitem[{{Kepley} {et~al.}(2020){Kepley}, {Tsutsumi}, {Brogan}, {Indebetouw},
  {Yoon}, {Mason}, \& {Donovan Meyer}}]{Kepley20}
{Kepley}, A.~A., {Tsutsumi}, T., {Brogan}, C.~L., {et~al.} 2020, \pasp, 132,
  024505, \dodoi{10.1088/1538-3873/ab5e14}

\bibitem[{{Kirk} {et~al.}(2013){Kirk}, {Myers}, {Bourke}, {Gutermuth},
  {Hedden}, \& {Wilson}}]{kirk2013}
{Kirk}, H., {Myers}, P.~C., {Bourke}, T.~L., {et~al.} 2013, \apj, 766, 115,
  \dodoi{10.1088/0004-637X/766/2/115}

\bibitem[{{Koch} \& {Rosolowsky}(2015)}]{koch15}
{Koch}, E.~W., \& {Rosolowsky}, E.~W. 2015, \mnras, 452, 3435,
  \dodoi{10.1093/mnras/stv1521}

\bibitem[{{Kumar} {et~al.}(2020){Kumar}, {Palmeirim}, {Arzoumanian}, \&
  {Inutsuka}}]{Kumar20}
{Kumar}, M.~S.~N., {Palmeirim}, P., {Arzoumanian}, D., \& {Inutsuka}, S.~I.
  2020, \aap, 642, A87, \dodoi{10.1051/0004-6361/202038232}

\bibitem[{{Larson}(1982)}]{larson1982}
{Larson}, R.~B. 1982, \mnras, 200, 159, \dodoi{10.1093/mnras/200.2.159}

\bibitem[{{Liu} {et~al.}(2012){Liu}, {Jim{\'e}nez-Serra}, {Ho}, {Chen},
  {Zhang}, \& {Li}}]{liu2012}
{Liu}, H.~B., {Jim{\'e}nez-Serra}, I., {Ho}, P. T.~P., {et~al.} 2012, \apj,
  756, 10, \dodoi{10.1088/0004-637X/756/1/10}

\bibitem[{{Loren}(1976)}]{loren76}
{Loren}, R.~B. 1976, \apj, 209, 466, \dodoi{10.1086/154741}

\bibitem[{{Matthews} {et~al.}(2014){Matthews}, {Ade}, {Angil{\`e}}, {Benton},
  {Chapin}, {Chapman}, {Devlin}, {Fissel}, {Fukui}, {Gandilo}, {Gundersen},
  {Hargrave}, {Klein}, {Korotkov}, {Moncelsi}, {Mroczkowski}, {Netterfield},
  {Novak}, {Nutter}, {Olmi}, {Pascale}, {Poidevin}, {Savini}, {Scott},
  {Shariff}, {Soler}, {Tachihara}, {Thomas}, {Truch}, {Tucker}, {Tucker}, \&
  {Ward-Thompson}}]{matthews2014}
{Matthews}, T.~G., {Ade}, P. A.~R., {Angil{\`e}}, F.~E., {et~al.} 2014, \apj,
  784, 116, \dodoi{10.1088/0004-637X/784/2/116}

\bibitem[{{Mizuno} {et~al.}(1995){Mizuno}, {Onishi}, {Yonekura}, {Nagahama},
  {Ogawa}, \& {Fukui}}]{mizuno95}
{Mizuno}, A., {Onishi}, T., {Yonekura}, Y., {et~al.} 1995, \apjl, 445, L161,
  \dodoi{10.1086/187914}

\bibitem[{{Moeckel} \& {Burkert}(2015)}]{moeckel2015}
{Moeckel}, N., \& {Burkert}, A. 2015, \apj, 807, 67,
  \dodoi{10.1088/0004-637X/807/1/67}

\bibitem[{{Muraoka} {et~al.}(2020){Muraoka}, {Kondo}, {Tokuda}, {Nishimura},
  {Miura}, {Onodera}, {Kuno}, {Zahorecz}, {Tsuge}, {Sano}, {Fujita}, {Onishi},
  {Saigo}, {Tachihara}, {Fukui}, \& {Kawamura}}]{Muraoka20}
{Muraoka}, K., {Kondo}, H., {Tokuda}, K., {et~al.} 2020, \apj, 903, 94,
  \dodoi{10.3847/1538-4357/abb822}

\bibitem[{{Muraoka} {et~al.}(2023){Muraoka}, {Konishi}, {Tokuda}, {Kondo},
  {Miura}, {Tosaki}, {Onodera}, {Kuno}, {Kobayashi}, {Tsuge}, {Sano}, {Kitano},
  {Fujita}, {Nishimura}, {Onishi}, {Saigo}, {Yamada}, {Demachi}, {Tachihara},
  {Fukui}, {Kawamura}, \& {AAS Journals Data Editors}}]{Muraoka23}
{Muraoka}, K., {Konishi}, A., {Tokuda}, K., {et~al.} 2023, \apj, 953, 164,
  \dodoi{10.3847/1538-4357/ace4bd}

\bibitem[{{Myers}(2009)}]{myers09}
{Myers}, P.~C. 2009, \apj, 700, 1609, \dodoi{10.1088/0004-637X/700/2/1609}

\bibitem[{{Nagahama} {et~al.}(1998){Nagahama}, {Mizuno}, {Ogawa}, \&
  {Fukui}}]{nagahama98}
{Nagahama}, T., {Mizuno}, A., {Ogawa}, H., \& {Fukui}, Y. 1998, \aj, 116, 336,
  \dodoi{10.1086/300392}

\bibitem[{{Nakamura} {et~al.}(2014){Nakamura}, {Sugitani}, {Tanaka},
  {Nishitani}, {Dobashi}, {Shimoikura}, {Shimajiri}, {Kawabe}, {Yonekura},
  {Mizuno}, {Kimura}, {Tokuda}, {Kozu}, {Okada}, {Hasegawa}, {Ogawa}, {Kameno},
  {Shinnaga}, {Momose}, {Nakajima}, {Onishi}, {Maezawa}, {Hirota}, {Takano},
  {Iono}, {Kuno}, \& {Yamamoto}}]{nakamura2014}
{Nakamura}, F., {Sugitani}, K., {Tanaka}, T., {et~al.} 2014, \apjl, 791, L23,
  \dodoi{10.1088/2041-8205/791/2/L23}

\bibitem[{{Neelamkodan} {et~al.}(2021){Neelamkodan}, {Tokuda}, {Barman},
  {Kondo}, {Sano}, \& {Onishi}}]{Neelamkodan21}
{Neelamkodan}, N., {Tokuda}, K., {Barman}, S., {et~al.} 2021, \apjl, 908, L43,
  \dodoi{10.3847/2041-8213/abdebb}

\bibitem[{{Neuhauser} \& {Brandner}(1998)}]{neuhauser1998}
{Neuhauser}, R., \& {Brandner}, W. 1998, \aap, 330, L29,
  \dodoi{10.48550/arXiv.astro-ph/9712045}

\bibitem[{{Neuh{\"a}user} \& {Forbrich}(2008)}]{neuhauser08}
{Neuh{\"a}user}, R., \& {Forbrich}, J. 2008, in Handbook of Star Forming
  Regions, Volume II, ed. B.~{Reipurth}, Vol.~5, 735,
  \dodoi{10.48550/arXiv.0808.3374}

\bibitem[{{Nutter} {et~al.}(2005){Nutter}, {Ward-Thompson}, \&
  {Andr{\'e}}}]{nutter2005}
{Nutter}, D.~J., {Ward-Thompson}, D., \& {Andr{\'e}}, P. 2005, \mnras, 357,
  975, \dodoi{10.1111/j.1365-2966.2005.08711.x}

\bibitem[{{Oasa} {et~al.}(2008){Oasa}, {Tamura}, {Sunada}, \&
  {Sugitani}}]{oasa2008}
{Oasa}, Y., {Tamura}, M., {Sunada}, K., \& {Sugitani}, K. 2008, \aj, 136, 1372,
  \dodoi{10.1088/0004-6256/136/3/1372}

\bibitem[{{Ostriker}(1964)}]{ostriker64}
{Ostriker}, J. 1964, \apj, 140, 1056, \dodoi{10.1086/148005}

\bibitem[{{Palmeirim} {et~al.}(2013){Palmeirim}, {Andr{\'e}}, {Kirk},
  {Ward-Thompson}, {Arzoumanian}, {K{\"o}nyves}, {Didelon}, {Schneider},
  {Benedettini}, {Bontemps}, {Di Francesco}, {Elia}, {Griffin}, {Hennemann},
  {Hill}, {Martin}, {Men'shchikov}, {Molinari}, {Motte}, {Nguyen Luong},
  {Nutter}, {Peretto}, {Pezzuto}, {Roy}, {Rygl}, {Spinoglio}, \&
  {White}}]{palmeirim13}
{Palmeirim}, P., {Andr{\'e}}, P., {Kirk}, J., {et~al.} 2013, \aap, 550, A38,
  \dodoi{10.1051/0004-6361/201220500}

\bibitem[{{Peretto} {et~al.}(2014){Peretto}, {Fuller}, {Andr{\'e}},
  {Arzoumanian}, {Rivilla}, {Bardeau}, {Duarte Puertas}, {Guzman Fernandez},
  {Lenfestey}, {Li}, {Olguin}, {R{\"o}ck}, {de Villiers}, \&
  {Williams}}]{peretto14}
{Peretto}, N., {Fuller}, G.~A., {Andr{\'e}}, P., {et~al.} 2014, \aap, 561, A83,
  \dodoi{10.1051/0004-6361/201322172}

\bibitem[{{Peterson} {et~al.}(2011){Peterson}, {Caratti o Garatti}, {Bourke},
  {Forbrich}, {Gutermuth}, {J{\o}rgensen}, {Allen}, {Patten}, {Dunham},
  {Harvey}, {Mer{\'\i}n}, {Chapman}, {Cieza}, {Huard}, {Knez}, {Prager}, \&
  {Evans}}]{peterson11}
{Peterson}, D.~E., {Caratti o Garatti}, A., {Bourke}, T.~L., {et~al.} 2011,
  \apjs, 194, 43, \dodoi{10.1088/0067-0049/194/2/43}

\bibitem[{{Pineda} {et~al.}(2023){Pineda}, {Arzoumanian}, {Andre}, {Friesen},
  {Zavagno}, {Clarke}, {Inoue}, {Chen}, {Lee}, {Soler}, \&
  {Kuffmeier}}]{pineda2023}
{Pineda}, J.~E., {Arzoumanian}, D., {Andre}, P., {et~al.} 2023, in Astronomical
  Society of the Pacific Conference Series, Vol. 534, Protostars and Planets
  VII, ed. S.~{Inutsuka}, Y.~{Aikawa}, T.~{Muto}, K.~{Tomida}, \& M.~{Tamura},
  233, \dodoi{10.48550/arXiv.2205.03935}

\bibitem[{{Ren} {et~al.}(2021){Ren}, {Zhu}, {Shi}, {Yue}, {Li}, {Zhang},
  {Mardones}, {Wu}, {Jiao}, {Liu}, {Luo}, {Xie}, {Zhang}, \& {Xu}}]{ren2021}
{Ren}, Z., {Zhu}, L., {Shi}, H., {et~al.} 2021, \mnras, 505, 5183,
  \dodoi{10.1093/mnras/stab1509}

\bibitem[{{Rossano}(1978)}]{rossano1978}
{Rossano}, G.~S. 1978, \aj, 83, 234, \dodoi{10.1086/112198}

\bibitem[{{Sato} {et~al.}(2023){Sato}, {Takahashi}, {Ishii}, {Ho}, {Machida},
  {Carpenter}, {A. Zapata}, {Teixeira}, \& {Suri}}]{sato2023}
{Sato}, A., {Takahashi}, S., {Ishii}, S., {et~al.} 2023, \apj, 944, 92,
  \dodoi{10.3847/1538-4357/aca7c9}

\bibitem[{{Sicilia-Aguilar} {et~al.}(2011){Sicilia-Aguilar}, {Henning},
  {Kainulainen}, \& {Roccatagliata}}]{sicilia-aguilar11}
{Sicilia-Aguilar}, A., {Henning}, T., {Kainulainen}, J., \& {Roccatagliata}, V.
  2011, \apj, 736, 137, \dodoi{10.1088/0004-637X/736/2/137}

\bibitem[{{Sicilia-Aguilar} {et~al.}(2013){Sicilia-Aguilar}, {Henning}, {Linz},
  {Andr{\'e}}, {Stutz}, {Eiroa}, \& {White}}]{sicilia-aguilar13}
{Sicilia-Aguilar}, A., {Henning}, T., {Linz}, H., {et~al.} 2013, \aap, 551,
  A34, \dodoi{10.1051/0004-6361/201220170}

\bibitem[{{Smith} {et~al.}(2016){Smith}, {Glover}, {Klessen}, \&
  {Fuller}}]{smith2016}
{Smith}, R.~J., {Glover}, S. C.~O., {Klessen}, R.~S., \& {Fuller}, G.~A. 2016,
  \mnras, 455, 3640, \dodoi{10.1093/mnras/stv2559}

\bibitem[{{Snow} \& {Hillier}(2021)}]{snow2021}
{Snow}, B., \& {Hillier}, A. 2021, \mnras, 506, 1334,
  \dodoi{10.1093/mnras/stab1672}

\bibitem[{{Tachihara} {et~al.}(2000){Tachihara}, {Abe}, {Onishi}, {Mizuno}, \&
  {Fukui}}]{tachihara2000}
{Tachihara}, K., {Abe}, R., {Onishi}, T., {Mizuno}, A., \& {Fukui}, Y. 2000,
  \pasj, 52, 1147, \dodoi{10.1093/pasj/52.6.1147}

\bibitem[{{Tachihara} {et~al.}(2002){Tachihara}, {Onishi}, {Mizuno}, \&
  {Fukui}}]{tachihara02}
{Tachihara}, K., {Onishi}, T., {Mizuno}, A., \& {Fukui}, Y. 2002, \aap, 385,
  909, \dodoi{10.1051/0004-6361:20020180}

\bibitem[{{Tokuda} {et~al.}(2023{\natexlab{a}}){Tokuda}, {Fukaya}, {Tachihara},
  {Omura}, {Harada}, {Nozaki}, {Shoshi}, \& {Machida}}]{Tokuda23Interc}
{Tokuda}, K., {Fukaya}, N., {Tachihara}, K., {et~al.} 2023{\natexlab{a}},
  \apjl, 956, L16, \dodoi{10.3847/2041-8213/acfca9}

\bibitem[{{Tokuda} {et~al.}(2019{\natexlab{a}}){Tokuda}, {Fukui}, {Harada},
  {Saigo}, {Tachihara}, {Tsuge}, {Inoue}, {Torii}, {Nishimura}, {Zahorecz},
  {Nayak}, {Meixner}, {Minamidani}, {Kawamura}, {Mizuno}, {Indebetouw},
  {Sewi{\l}o}, {Madden}, {Galametz}, {Lebouteiller}, {Chen}, \&
  {Onishi}}]{Tokuda19N159}
{Tokuda}, K., {Fukui}, Y., {Harada}, R., {et~al.} 2019{\natexlab{a}}, \apj,
  886, 15, \dodoi{10.3847/1538-4357/ab48ff}

\bibitem[{{Tokuda} {et~al.}(2019{\natexlab{b}}){Tokuda}, {Tachihara}, {Saigo},
  {Andr{\'e}}, {Miyamoto}, {Zahorecz}, {Inutsuka}, {Matsumoto}, {Takashima},
  {Machida}, {Tomida}, {Taniguchi}, {Fukui}, {Kawamura}, {Tatematsu},
  {Kandori}, \& {Onishi}}]{tokuda19MC5}
{Tokuda}, K., {Tachihara}, K., {Saigo}, K., {et~al.} 2019{\natexlab{b}}, \pasj,
  71, 73, \dodoi{10.1093/pasj/psz051}

\bibitem[{{Tokuda} {et~al.}(2020){Tokuda}, {Fujishiro}, {Tachihara},
  {Takashima}, {Fukui}, {Zahorecz}, {Saigo}, {Matsumoto}, {Tomida}, {Machida},
  {Inutsuka}, {Andr{\'e}}, {Kawamura}, \& {Onishi}}]{tokuda20}
{Tokuda}, K., {Fujishiro}, K., {Tachihara}, K., {et~al.} 2020, \apj, 899, 10,
  \dodoi{10.3847/1538-4357/ab9ca7}

\bibitem[{{Tokuda} {et~al.}(2021){Tokuda}, {Kondo}, {Ohno}, {Konishi}, {Sano},
  {Tsuge}, {Zahorecz}, {Goto}, {Neelamkodan}, {Wong}, {Sewi{\l}o}, {Fukushima},
  {Takekoshi}, {Muraoka}, {Kawamura}, {Tachihara}, {Fukui}, \&
  {Onishi}}]{Tokuda21}
{Tokuda}, K., {Kondo}, H., {Ohno}, T., {et~al.} 2021, \apj, 922, 171,
  \dodoi{10.3847/1538-4357/ac1ff4}

\bibitem[{{Tokuda} {et~al.}(2023{\natexlab{b}}){Tokuda}, {Harada}, {Tanaka},
  {Inoue}, {Shimonishi}, {Zhang}, {Sewi{\l}o}, {Kunitoshi}, {Konishi}, {Fukui},
  {Kawamura}, {Onishi}, \& {Machida}}]{Tokuda23LMC}
{Tokuda}, K., {Harada}, N., {Tanaka}, K. E.~I., {et~al.} 2023{\natexlab{b}},
  \apj, 955, 52, \dodoi{10.3847/1538-4357/acefb7}

\bibitem[{{Tokuda} {et~al.}(2024){Tokuda}, {Harada}, {Omura}, {Matsumoto},
  {Onishi}, {Saigo}, {Shoshi}, {Nozaki}, {Tachihara}, {Fukaya}, {Fukui},
  {Inutsuka}, \& {Machida}}]{tokuda2024}
{Tokuda}, K., {Harada}, N., {Omura}, M., {et~al.} 2024, arXiv e-prints,
  arXiv:2403.00305, \dodoi{10.48550/arXiv.2403.00305}

\bibitem[{{Yonekura} {et~al.}(1999){Yonekura}, {Mizuno}, {Saito}, {Mizuno},
  {Ogawa}, \& {Fukui}}]{yonekura99}
{Yonekura}, Y., {Mizuno}, N., {Saito}, H., {et~al.} 1999, \pasj, 51, 911,
  \dodoi{10.1093/pasj/51.6.911}

\end{thebibliography}
